\documentclass[twocolumn]{svjour3}          % twocolumn
\usepackage{amsmath}
\usepackage{amssymb}
\usepackage[utf8]{inputenc} % allow utf-8 input
\usepackage[T1]{fontenc}    % use 8-bit T1 fonts
\usepackage{hyperref}       % hyperlinks
\usepackage{url}            % simple URL typesetting
\usepackage{booktabs}       % professional-quality tables
\usepackage{amsfonts}       % blackboard math symbols
\usepackage{nicefrac}       % compact symbols for 1/2, etc.
\usepackage{microtype}      % microtypography
\usepackage{lipsum}
\usepackage{xcolor}
\usepackage{graphicx}
\usepackage{dsfont}
\usepackage[ruled,vlined]{algorithm2e}
\usepackage{csvsimple}
\usepackage{bbm}
\usepackage[numbers]{natbib}
\usepackage{multicol}
\usepackage{resizegather}
\usepackage{ulem}
\usepackage{caption,subcaption}

\begin{document}

\title{Group Testing Enables Asymptomatic Screening for COVID-19 Mitigation}
\subtitle{Feasibility and Optimal Pool Size Selection with Dilution Effects}

%\titlerunning{Short form of title}        % if too long for running head

\author{
Yifan Lin \and Yuxuan Ren \and Jingyuan Wan \and Massey Cashore \and Jiayue Wan \and Yujia Zhang \and Peter Frazier \and Enlu Zhou
}

% \authorrunning{Short form of author list} % if too long for running head

\institute{Yifan Lin \at
                Georgia Institute of Technology, Atlanta, GA 30332, USA
            \and
            Yuxuan Ren \at
                Georgia Institute of Technology, Atlanta, GA 30332, USA
            \and
            Jingyuan Wan \at
                Georgia Institute of Technology, Atlanta, GA 30332, USA
            \and
            Massey Cashore \at
                Cornell University, Ithaca, NY 14853, USA
            \and
            Jiayue Wan \at
                Cornell University, Ithaca, NY 14853, USA
            \and
            Yujia Zhang \at
                Cornell University, Ithaca, NY 14853, USA
            \and
            Peter Frazier \at
                Cornell University, Ithaca, NY 14853, USA\\ \email{pf98@cornell.edu}
            \and 
            Enlu Zhou \at
                Georgia Institute of Technology, Atlanta, GA 30332, USA\\\email{enlu.zhou@isye.gatech.edu}
}

% \date{Received: date / Accepted: date}
% The correct dates will be entered by the editor
\maketitle

\begin{abstract}
Repeated asymptomatic screening for SARS-CoV-2 promises to control spread of the virus but would require too many resources to implement at scale. 
Group testing is promising for screening more people with fewer test resources: multiple samples tested together in one pool can be excluded with one negative test result. 
Existing approaches to group testing design for SARS-CoV-2 asymptomatic screening, however, do not consider dilution effects: that false negatives become more common with larger pools.
As a consequence, they may recommend pool sizes that are too large or misestimate the benefits of screening.
Modeling dilution effects,
we derive closed-form expressions for the expected number of tests and false negative/positives per person screened under two popular group testing methods: the linear and square array methods.  
We find that test error correlation induced by a common viral load across an individual's samples results in many fewer false negatives than would be expected from less realistic but more widely assumed independent errors.
This insight also suggests that false positives can be controlled through repeated tests without significantly increasing false negatives.  
Using these closed-form expressions to trace a Pareto frontier over error rates and tests,
we design testing protocols for repeated asymptomatic screening of a large population. 
We minimize disease prevalence by optimizing a time-varying pool sizes and screening frequency constrained by daily test capacity and a false positive limit.
This provides a testing protocol practitioners can use for mitigating COVID-19.
In a case study, we demonstrate the effectiveness of this methodology in controlling spread. 

\keywords{COVID-19 \and Group testing \and Dilution effect \and Asymptomatic screening \and Prevalence control}
\end{abstract}

\section*{Highlights}
\begin{itemize}

\item We demonstrate the feasibility of group testing for enabling population-level asymptomatic screening for SARS-CoV-2, including practically significant dilution effects in our analysis. 

\item In a case study with a nominal set of parameters, we demonstrate that screening a population of 100K people twice per week with square array group testing and a pool size near 25 can keep prevalence from growing with approximately 20K tests / week. At the same time, not screening allows prevalence to double every 4 days.

\item Our analysis shows that, compared to the linear array method, the square array method is more feasible  when the testing capacity is limited, and generally has slightly higher (but comparable) false negative rates and much lower false positive rates. Therefore, we recommend using the square array group testing method.

\item We provide practitioners with an implementable protocol, including how to determine the optimal cycle length in repeated asymptotic screening and the optimal pool size within each testing cycle, and the procedure of group testing.

\item A common concern with group testing is that multiple tests with independent errors prevent achieving the high sensitivity. We show that false negatives driven by low viral load are positively correlated across pools resulting in much higher sensitivity than obtained from a naive analysis. At the same time, false positive pools driven independently by lab contamination result in a low overall false positive rate (high specificity).
\end{itemize}

\section{Introduction}
\label{intro}
Repeated asymptomatic screening of an entire population promises to effectively combat COVID-19 \citep{gollier2020group,paltiel2020,chang2020,chen2020,peter2020}. 
Asymptomatic screening identifies infectious but asymptomatic individuals who can then be isolated. 
These individuals can also initiate contact tracing to isolate more infectious individuals.
Asymptomatic screening is particularly attractive as an intervention for COVID-19 because so transmission by asymptomatic and presymptomatic individuals plays such a large role in its spread.
Screening repeatedly allows false negatives missed in one round of screening to be identified in a subsequent round. It also protects against new infections imported to a population through travel. 
By screening more frequently, we can reduce the time period over which an infectious individual can transmit virus. 
If done often enough to a large-enough fraction of a self-contained population, then both mathematical modeling 
\citep{gollier2020group,paltiel2020,chang2020,chen2020,peter2020}
and real-world experience 
\citep{bloomberg2020}
suggest that repeated asymptomatic screening and taking appropriate measures (such as quarantine) can drive the prevalence of COVID-19 to almost 0, even if the level of social distancing and mask-wearing are not enough to prevent spread by themselves.

For example, \cite{paltiel2020} found that frequent asymptomatic screening along with behavioral interventions can keep infections on a college campus under control and permit safe reopening; \cite{chang2020} used a similar approach as \cite{paltiel2020} to evaluate different screening frequencies, while accounting for temporal variations in infectivity and test detectability of positive cases; \cite{chen2020} used an SIR model subject to testing capacity constraints to evaluate the effect of testing capacity and mitigation policies on curbing the spread of the epidemic in a general population; \cite{bloomberg2020} reports on the success of a repeated asymptomatic screening program at Cornell University.

However, the resources necessary to screen a large population as frequently as twice per week would seem to be prohibitive. 
Group testing \cite{dorfman1944} is a promising approach for overcoming this challenge.
In the simplest version of this approach as it is applied to SARS-CoV-2, Dorfman's procedure or the so-called linear array method, samples are first collected from all individuals we wish to screen. Then, a portion from each of multiple individuals' samples are taken, combined together in a single "pool", and the pool is tested in a single chemical reaction (typically PCR, or polymerase chain reaction) for the presence of the virus. If the pool tests negative, then (in the absence of errors) all of the individuals participating in the pool can be determined to be virus-free.
If the pool tests positive, then additional tests are performed on each individual sample to determine which ones contain virus. In low-prevalence populations, this approach can significantly reduce the number of PCR reactions that need to be done, saving chemical reagents, machine time, and the time of the trained laboratory personnel needed to supervise these reactions.
(Notably, group testing does not save on the effort and resources required to collect the samples themselves.)

Building on earlier success in using group testing to conserve
resources when screening for malaria, HIV, and H1N1 \citep{hsiang2010pcr,pilcher2002real,kim2007comparison,westreich2008optimizing,van2012pooling},
and demonstrations that it could provide similar resource savings when testing for SARS-CoV-2 
\citep{Mentus2020,broder2020note,armendriz2020group,stefan2020, narayanan2020accelerated,eberhardt2020multi,bateman2020assessing},
there has been substantial interest in the use of group testing with population-level asymptomatic screening to control the spread of the virus.

Indeed, several recent mathematical analyses have advocated for this.
For example, \cite{gollier2020group} found the optimal pool size that released as many individuals to the workforce as possible without considering error in group testing. The expanded test capacity further enables frequent asymptomatic screening for a large community within a short period of time. \cite{peter2020} simulated using group testing to monitor and control the prevalence on Cornell's campus, and \cite{mutesa2020} devised a hypercube-array pooling algorithm that enabled resource-efficient screening in Rwanda.

All of the aforementioned literature studying group testing and asymptomatic screening for control of COVID-19 assumes either that testing is error-free, or the false negative rate is a fixed constant, regardless of pool size.
In reality, the viral load in virus-containing samples (referred to as positive samples in the rest of this paper) is diluted when lower volume is used in the reaction to support pooling with negative samples. This lowers their probability of being detected and is called the {\it dilution effect}.
As a consequence of not modeling errors or modeling them as a small constant appropriate for small pool sizes, the previous literature risks overstating the benefits of group testing and risks recommending protocols with unrealistically large pool sizes. At the same time, if one uses a constant error probability appropriate for a large pool size, one can underestimate the benefits and fail to take advantage of the ability to reduce errors by using smaller pools.  
Indeed, this inattention to the dilution effect is reflective of the larger group testing literature, in which only a handful of papers \cite{wein1996pooled,zenios1998pooled, pilcher2020group,cleary2020using} consider this effect in any group testing design context and none that we are aware of in the context of designing group testing protocols for prevalence control with repeated asymptomatic screening.
(\cite{pilcher2020group} designs test protocols to maximally expand the screening capacity under different prevalence levels at some sacrifice of sensitivity, and the concurrent-to-our work \cite{cleary2020using} designs group testing protocols to maximize the number of true positives found, both without quantitatively studying its ability to control prevalence through repeated application.)  \

In this paper, we consider dilution effects and the choice of pool sizes in group testing protocols for specific use in asymptomatic screening for controlling the prevalence of COVID-19. 
We first quantify the trade-off between the number of pooled tests performed, the number of false negatives and the number of false positives. 
In particular, under dilution effects we derive closed-form expressions of these quantities 
as a function of pool size in the context of two established and practical group testing protocols, Dorfman's linear array and the square array method.

A notable outcome of our model of dilution effects is that false positives can be kept small while false negatives can be prevented from becoming too large. A naive analysis of group testing with either the linear or square array method is that false negatives and false positives in pooled tests occur independently across tests. While one might wish to reduce false positives (which is very important in asymptomatic screening) by testing a sample one additional time, a belief in independence of false negatives would suggest that this would also increase the false negative rate. This could be a significant enough issue to make group-testing-based asymptomatic screening infeasible. We argue, however, that correlation in false negatives induced by similar viral loads in portions of the same sample used in multiple pooled tests allows the use of repeated tests to control false positives without significantly increasing the false negative rate. 

Using our model of dilution effects, we then compute the optimal pool size that minimizes the expected number of false negatives per person while constraining test capacity and false positives per person. 

This formulation is particularly appropriate for asymptomatic screening when the number of pooled tests that can be performed in a day is a key limiting factor. This is imposed when the number of machines that can perform these tests is limited or when the number of trained personnel available to operate them is limited. The limit on false positives per person tested is additionally imposed by a need to avoid quarantining uninfected people and the fact that frequent asymptomatic screening would test large numbers of people every day.
By varying these constraints one can trace out a Pareto frontier between expected pooled tests, false negatives, and false positives per person screened.

Using this more realistic model with dilution effects and the results of our optimization approach, we further consider repeated asymptomatic screening for a large population to control the prevalence of COVID-19 with constraints on pooled test capacity and false positives. 
To model the dynamics of virus spread, we propose a testing-quarantine-infection model that builds on SIR models from the epidemiology literature, divides the population into a chosen number of subsets (according to the cycle length) and then tests one subset on each day, testing that subset again in the next testing cycle. We fix the pool size for practicality during each cycle, but allow it to change as each cycle passes because prevalence may change significantly.

Assuming that the false positive rate is already constrained to be small enough to be negligible (enabled by our insight that it can be driven to $0$ through an additional test on those that would otherwise be positive), an asymptomatic screening policy should use a group testing policy that operates on the Pareto frontier between false negatives and tests per person screened: achieving a lower false negative rate results in fewer infections; using a lower number of tests per person screened allows screening more people per day which allows more frequent screening which results in fewer infections.

Leveraging the ability to constrain our search over pool sizes using the Pareto frontier significantly improves performance.
We compute by optimization via simulation the optimal testing cycle length that yields the minimal final prevalence rate after a given period of time subject to constraints on test capacity and false positives per person tested.
Our framework can also be extended to incorporate constraints on the total number of people tested, as would arise from limited sampling supplies such as swabs, tubes, viral transport media, or personnel needed to collect samples. 

Leveraging this methodology, we study the feasibility of group testing for enabling population-level asymptomatic screening that prevents the growth of COVID-19 outbreaks. We consider three scenarios varying initial population prevalence, transmission rates, and other parameters seeking to control prevalence in a population of 100K people using 40K pooled tests in total over 2 weeks. In a pessimistic scenario we find, unfortunately, that the rate of transmission is too high to be contained with this many tests.  In nominal and optimistic scenarios, however, we find that asymptomatic screening enabled by group testing is able to control and even reduce prevalence effectively in spite of the dilution effect. Our approach is able to respond to the dilution effect to recommend smaller pool sizes than those often be recommended by models that do not consider it \citep{peter2020}: optimal pool sizes chosen range from 10 to 35.

In contrast with other work on asymptomatic screening with group testing that does not consider sampling errors, such as \cite{gollier2020group}, the inclusion of sampling errors necessitates a different optimization framework. While 
\cite{gollier2020group} minimizes the expected number of pooled tests performed, the inclusion of errors necessitates simultaneous consideration of the trade-off between pooled tests performed, false negatives and false positives.

Our study of dilution effects is most closely related to \cite{wein1996pooled}, which considered dilution effects in group testing for HIV screening (using an ELISA assay rather our focus on PCR) and derived pool sizes and protocols that minimize a linear combination of the number of pooled tests, false negatives, and false positives.
We differ, however, in several important ways.
First, we differ in our focus on a PCR assay for COVID-19.
Second, we differ in our focus on simpler and more easily implemented linear and square array group testing protocols, rather than more complex multi-stage adaptive designs.
This derives in part from the need in asymptomatic screening for COVID-19 to report results quickly so that infectious individuals can be identified before they spread virus to more people. 
In contrast, multi-stage adaptive group testing, in which the next pool to test depends on the result of previous one, incurs delays with each additional stage.
The non-adaptive group testing methods we use are comparable in performance to adaptive methods such as generalized binary search (GBS) (e.g. \cite{theagarajan2020group}. 
Third and finally, we differ in the fact that we focus on the use of group testing specifically within asymptomatic screening for outbreak prevention.
This necessitates differences such as the difference in the optimization problem formulation. Rather than optimizing a linear combination of metrics we optimize false negatives subject to constraints on pooled tests and false positives. This enables tracing a full Pareto frontier rather than just the convex envelope that could be traced by varying weights of a linear objective. 

In summary, our contributions are three-fold:
\begin{enumerate}
    \item We develop models at the molecular level, testing level, and population level. The molecular-level model considers the viral load and false negative rate in pooled samples. The testing-level model details the linear array and square array group testing methods. The population-level model captures the epidemic dynamics of testing, quarantine, and infection over time.  
    \item We derive closed-form expressions for the expected number of tests, the expected number of false negatives and the expected number of false positives for the two group testing methods. We compute the optimal pool size that minimizes the expected number of false negatives under test capacity and false positive constraints in a stochastic optimization formulation.
    \item We design a testing protocol to test the whole population in a testing cycle under a limited daily testing capacity. We further conduct a case study to demonstrate the testing protocol in different scenarios.
\end{enumerate}

While we view our analysis as significantly expanding our understanding of whether group testing can reduce resource requirements significantly enough to enable repeated asymptomatic screening at scale for control of COVID-19, our work has several limitations.
First, our model of disease progression in the general population uses simple stylized dynamics that do not include social network structure, dispersion in transmission, or time-varying infectivity that may significantly impact virus spread.
Second, while our model of test error includes viral load and its variability across individuals, it does not model how viral load varies systematically over the course of a person's infection as in \cite{pilcher2020group} and how this variation could interact with screening.  Temporal variation in the ability to detect infections has been shown to impact screening's ability to control prevalence in \cite{chang2020}.
Third and finally, it does not model uncertainty about prevalence and transmission and how this may impact the choice of group size (as in \cite{pilcher2020group}) and screening frequency (as in \cite{peter2020}).
These limitations should be kept in mind when interpreting our results.

The remainder of the paper is organized as follows. Section \ref{sec:LitReview} reviews the literature on group testing. Section \ref{sec:Dilution} describes our molecular-level model of the dilution effect, consisting of viral load in positive samples, the mechanism of a RT-qPCR test, and false negatives in pooled samples. Section \ref{sec:GroupTestingMethods} details the linear array and square array methods, derives their expected number of tests and expected number of false negatives/positives per person, explains why correlation from viral load controls error rates, and computes the optimal pool size. Section \ref{sec:TestingCycle} considers prevalence control at the population level, proposes a testing-quarantine-infection model, finds the optimal testing cycle length and the optimal pool size within each cycle. We end with a case study of monitoring and controlling the prevalence in a large population under different scenarios in Section \ref{sec:CaseStudy}.

An early version of this paper circulated as a working paper \cite{frazier2020}. The current version is refined and expanded.

\section{Literature Review}
\label{sec:LitReview}
This section reviews the literature on group testing, screening for disease control, false negatives in group testing and in particular, errors from dilution effects.
\subsection{Group Testing: Methods and Applications}
\label{subsec:MethodsApp}
Group testing refers to the idea of pooling multiple samples together for testing to identify positives more efficiently. It was first studied by Dorfman in 1943 in the context of screening soldiers for syphilis during WWII \cite{dorfman1944}. 
Building on Dorfman's procedure, \cite{phatarfod1994use} proposed the square array method in 1994, which was closely analyzed later in \cite{westreich2008optimizing} in 2008. Under this approach, samples are placed onto a square array where each row and column forms a pool. Samples whose row and column pools both test positive are either deemed positive or tested in follow-up individual tests. \cite{Sobel1959} further generalized \cite{dorfman1944} to allow choosing pool sizes adaptively based on information from previous test results. These protocols have been generalized and analyzed using connections to 
active learning,
coding theory 
and compressed sensing \citep{Mentus2020, Atia2012,Malioutov2012, indyk2010efficiently,Malyutov2013,Aldridge2019Group}.

The need for detecting infectious diseases such as HIV, H1N1, and malaria has motivated the development of practically feasible group testing methods. \cite{pilcher2002real} devised a multistage pooling method for the detection of HIV. Samples are put into intermediate pools of size ten which themselves form master pools. An individual sample is retested only if their parent-level pool tests positive. \cite{kim2007comparison} studied the efficiency and error rates of square-array method for detecting HIV infections and \cite{westreich2008optimizing} extended this analysis to multi-stage linear arrays. The feasibility of group testing has also been confirmed by experimental studies. \cite{hsiang2010pcr} examined the hierarchical linear array pooling method for the detection of malaria subject to experiment-level constraints. \cite{van2012pooling} studied the practicality of pooling throat swab specimens for detecting H1N1, discovering that forming pools of size ten does not reduce sensitivity. The general conclusion we can draw from the literature is that, compared to individual testing, group testing can reduce the overall number of tests needed and increase testing throughput.

\subsection{Asymptomatic Screening Using Group Testing}

In combating infectious diseases, group testing is particularly useful since it alleviates the resource challenges on conducting large-scale asymptomatic screening of the population. For many past infectious diseases, screening has been shown to be crucial for identifying infected individuals and preventing further spread \citep{cowling2010entry,kuehnert2016screening,brown2007routine}. Large-scale screening is especially useful when carriers are often asymptomatic, as with COVID-19 \citep{bai2020presumed}. Modeling studies have indicated the efficacy of screening for controlling COVID-19. 
Both \cite{paltiel2020} and \cite{chang2020} found that frequent asymptomatic screening is essential for keeping infections on a college campus under control; \cite{chen2020} found that expanding testing capacity for large-scale screening, along with other mitigation policies, can effectively help curb the spread in a general population. 

The possibility of using group testing to conduct screening for COVID-19 has been studied both methodologically and empirically. Methodology-wise, \cite{Mentus2020} proposed both non-adaptive and adaptive group testing strategies based on generalized binary splitting. 
\cite{broder2020note} proposed a double pooling strategy and derived the optimal group size that minimizes the expected number of tests per person. \cite{armendriz2020group}, \cite{narayanan2020accelerated}, and \cite{eberhardt2020multi} examined mathematical properties of multi-stage pooling methods that generalize Dorfman's procedure. Empirically, \cite{shental2020efficient} designed a combinatorial pooling strategy based on Reed-Solomon error correcting codes improving test efficiency by a factor of 8 in a screening pilot. \cite{ben2020large} implemented Dorfman's procedure for screening more than 26,000 individuals and achieved a 7.3-fold increase in throughput. Our work builds on this promise but goes beyond past work:
except for a few examples discussed in detail below, existing work in group testing for asymptomatic screening either ignores important real-world aspects of test errors or does not fully model the interaction of group testing properties with the needs and dynamics of repeated asymptomatic screening for prevalence control.

\subsection{False Negatives and Dilution in Group Testing}
\label{subsec:FalseNeg}

While expanding the test capacity makes large-scale screening possible, ensuring test accuracy is also crucial for correctly identifying the positive cases and containing the spread. Accounting for the false negatives that arise in group testing is thus an essential part of the modeling. 

Group testing where the false negative rate is assumed to be a fixed constant has been studied theoretically. \cite{Graff1972GroupTI} and \cite{Graff1974} extended Dorfman's procedure \cite{dorfman1944} and Sobel and Groll's procedure \cite{Sobel1959} to the case where test error is present. \cite{KNILL1998} provided a unified framework that accounts for errors due to variable detectability of the samples, pooling errors, and screening errors, and showed how these factors affect the design of non-adaptive group testing protocols. In the associated area of error-correcting codes, samples are assigned to multiple pools combinatorially such that results can be robustly decoded under fixed false negative and false positive rates. \cite{cheraghchi2009compressed} discussed two decoding procedures for controlling the number of pooled tests used to detect all infected individuals under false positive probability. \cite{barg2017group} also focused on the decoding procedure and developed connections between group testing schemes and several error-correcting codes, such as Reed-Solomon code and algebraic-geometric code. 

In practice, the false negative rate is not constant. Rather, it depends on the pool size due to the dilution effect, i.e., when multiple samples are pooled into one sample, a virus-containing sample is effectively diluted among other negative samples, leading to false negatives if the combined viral load in the pooled sample is too low \cite{wein1996pooled,westreich2008optimizing}.
\cite{hwang1976} first proposed a conceptual model for dilution effects in pooled tests. \cite{weusten2011} proposed a probit model for the sensitivity of pooled test for HIV assuming that the pool contains no more than one positive sample. \cite{nguyen2019} expanded this probit model to account for temporal variation in test detectability of positive samples and allowed multiple positive samples in a pool. \cite{Claudia2006} used a probabilistic model to evaluate the dilution effect in the detection of bovine viral diarrhea virus. With regard to the recent COVID-19 pandemic, \cite{cleary2020using} used simulations to study the dilution effect in pooled tests while considering heterogeneity in viral progression across a large population. \cite{pilcher2020group} developed a similar temporal viral progression model and evaluated different pooling strategies (2-stage and 3-stage linear arrays) under different prevalence levels with dilution effect directly proportional to pool size. \cite{theagarajan2020group} applied the model developed by \cite{nguyen2019} to COVID group testing. Based on the mechanism of RT-qPCR tests and the clinical data for SARS-CoV-2, \cite{brault2020group} proposed a statistical model for determining the false negative rate induced by pooling. Recently, experimental studies on pooled SARS-CoV-2 tests have shown that pooling up to 30 samples is feasible with reasonable detection accuracy \cite{stefan2020} and \cite{yelin2020evaluation}.  

Among the aforementioned literature, \cite{brault2020group} is the closest to our molecular-level model of false negatives in PCR tests. However, we capture the physical properties of PCR tests in a more realistic way by modeling the probability of a positive test result as varying smoothly in the viral load and vanishing in the limit as the viral load goes to 0.
\section{Modeling Dilution Effects in Group Testing}
\label{sec:Dilution}
In this section, we present a molecular-level model of the false negative rate for a single pooled test as a function of sample viral load. This molecular-level model takes into account the dilution effect in the pooled test.

\subsection{PCR Preliminaries}
\label{subsec:PCRPre}
Real-time reverse transcription polymerase chain reaction (RT-qPCR) testing is the standard technique used to measure the amount of a specific RNA sequence in a sample. RNA copies in the sample are first reverse transcribed into complementary DNA sequences (cDNAs), which then get amplified through a polymerase chain reaction (PCR) in a PCR machine. In each cycle, the number of cDNA copies is approximately doubled. At the end of the test, the cycle threshold value ($C_t$) is returned, which indicates the number of cycles required for the cDNA concentration in the sample to reach a fluorescence-detectable threshold. Hence, the $C_t$ value is approximately equal to $-\log_2 V$ up to an additive constant, where $V$ is the viral RNA concentration (i.e. viral load) in the sample. The limit of detection (LoD) of the PCR test is the lowest viral load in the sample that can be detected in a PCR test with a specified probability (typically 95\%).

Possible sources of false negatives in individual RNA-based tests include the following:
\begin{itemize}
    \item Sample collection. Samples collected from infected individuals using swab (nasopharyngeal or oropharyngeal) or saliva may fail to contain enough amount of viral material due to improper sample collection and handling.
    \item Low viral load. The viral load in the individual body may be low, hence the viral load in the collected sample may be below the LoD of the PCR test.
\end{itemize}

In pooled RNA-based tests, the dilution effect is a major source of error. The dilution effect can be intuitively understood as dilution of viral load in the pooled sample that reduces test sensitivity. Modeling the dilution effect accurately relies on sampling from a realistic distribution of individual viral loads, which we discuss below in Section \ref{subsec:viral_load_dist}.

\subsection{Viral Load Distribution Among Infected Individuals} \label{subsec:viral_load_dist}

Based on data on 3,303 positive cases from large-scale screening in Germany \cite{jones2020}, \cite{brault2020group} established a Gaussian mixture model for the distribution of $C_t$ value of a single positive sample collected from an infected individual:
\begin{gather}
f(c) = \sum\limits_{k=1}^{3}\pi_k N(c; \mu_k, \sigma_k),
\label{eq:dist}
\end{gather}
where the parameters are estimated to be:
\begin{gather*}
\pi = [0.32, 0.54, 0.14], \mu = [20.14, 29.35, 34.78], \sigma = [3.60, 2.96, 1.32].
\end{gather*}

Note that this effectively fits a log Gaussian mixture model on the viral load among infected individuals. 

\subsection{Fitting the False Negative Rate Model} \label{subsec:fnr}
\cite{brault2020group} estimated the $C_t$ value corresponding to the LoD of the RT-qPCR test used in their data to be $d_{cens} = 35.6$. They suppose that a test returning a $C_t$ value less than $d_{cens}$ will deem the sample to be positive; otherwise, they suppose that the sample is deemed positive with probability $\beta = 0.2$. 

This model is simple, yet it fails to capture an essential feature of PCR tests. In fact, the false negative rate is expected to strictly decrease with the sample viral load. Once the viral load falls below a certain threshold, the false negative rate is expected to quickly approach zero. This pattern can be well represented by a logistic function  (e.g. \cite{forootan2017methods}). In light of this, we fit a logistic curve to the relationship between the $C_t$ value (i.e., negative log viral load) and the false negative rate (FNR)
\begin{gather}
    FNR(c)=\frac{1}{1+e^{-k(c-c_0)}},
    \label{eq:logistic}
\end{gather}
so that it is roughly consistent with the model proposed in \cite{brault2020group}. Ideally, we would pick parameters $k$ and $c_0$ such that the curve $FNR(c)$ satisfies the following three conditions:
\begin{itemize}
    \item[(i)] It finds that $d_{cens}$ is the LoD. Since the LoD is conventionally defined as the lowest viral load that can be detected in a PCR test 95\% of the time, we would like $$FNR(d_{cens})=0.05.$$
    \item[(ii)] The expected false negative rate for samples beyond the LoD would match that of the model proposed in \cite{brault2020group}. Mathematically, we would like 
    \begin{align*}
        \mathbb{E}[FNR(c)\mid c>d_{cens}]&=\displaystyle\frac{\int_{d_{cens}}^{+\infty}FNR(c)f(c)dc}{\int_{d_{cens}}^{+\infty}f(c)dc}\\
        &=1-\beta.
    \end{align*}
    \item[(iii)] The expected overall FNR in the two error models match. Mathematically, we would like
    \begin{equation}
        \int_{-\infty}^{+\infty}FNR(c)f(c)dc = \int_{d_{cens}}^{+\infty}(1-\beta)f(c)dc.
    \label{eq:condition3}
    \end{equation}
\end{itemize}

\begin{figure}[htbp]
    \centering
    \includegraphics[width=.4\textwidth]{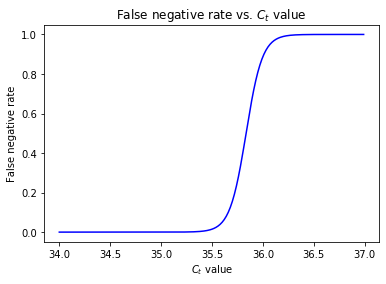}
    \caption{Fitted logistic curve for modeling the false negative rate as a function of the $C_t$ value.}
    \label{fig:fnr}
\end{figure}

Given that we have two parameters to fit in the logistic curve, all three conditions cannot be satisfied at the same time. We use conditions (i) and (ii) to fit the curve and obtain $\hat{k}=12.5, \hat{c}_0=35.8$ in equation (\ref{eq:logistic}). The fitted curve is shown in Figure \ref{fig:fnr}.

Plugging the fitted values of $k$ and $x_0$ into condition (iii), we find that the absolute difference between the LHS and RHS of equation (\ref{eq:condition3}) is only $4.6\times 10^{-4}$. Hence, we consider condition (iii) to be approximately satisfied.

\subsection{False Negative Rate of Pooled Samples}
\label{subsec:fnrPooled}
\begin{figure}[htb]
    \centering
    \includegraphics[width=.4\textwidth, ]{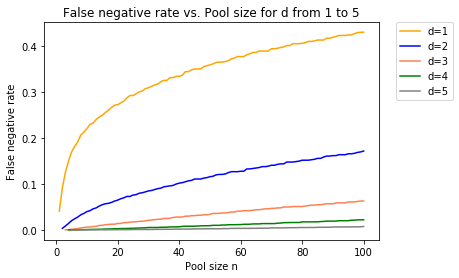}
    \caption{Simulated expected false negative rate $\gamma(n,d)$ vs. pool size $n$ under the presence of different numbers of positives $d$ in the pool.} 
    \label{fig: FN_matrix}
\end{figure}

In Section \ref{subsec:fnr}, we discussed the FNR model of a single positive sample. When group testing is implemented, however, in general we might expect a pooled sample to contain $n$ sub-samples collected from distinct individuals, with $d$ $(d\le n)$ positive samples among them. Let $\gamma(n,d)$ denote the expected false negative rate of a pooled sample of size $n$ that contains $d\geq 1$ positive sub-samples, where the expectation is taken over the randomness in individual viral loads. Let $V_i$, $i=1,2,\cdots, d$ be the viral load in the $d$ positive sub-samples. Assuming we take an equal portion of each sub-sample, the average viral load in the pooled sample is $\frac{1}{n}\sum_{i=1}^{d}V_i$, which results in a $C_t$ value of the pooled sample of 
\begin{gather}\label{eq:pooled_sample}
C_t = -\log_2\left(\frac{1}{n}\sum\limits_{i=1}^{d}V^i\right) = -\log_2\left(\frac{1}{n}\sum\limits_{i=1}^{d}2^{-C_{t,i}}\right),
\end{gather}
where $C_{t,i}$ is the $C_t$ value of the $i$th sub-sample if it were tested in an individual (i.e., non-diluted) PCR test. We estimate $\gamma(n,d)$ using Monte Carlo simulation. In each replication, we
\begin{itemize}
    \item generate $d$ i.i.d random variables $C_{t,i}$, $i=1,2,\cdots, d$ from the distribution specified in (\ref{eq:dist});
    \item compute the $C_t$ value of the pooled sample using equation (\ref{eq:pooled_sample});
    \item compute the false negative rate of the pooled sample using equation (\ref{eq:logistic}).
\end{itemize}
The resulting simulated false negative rate is shown in Figure \ref{fig: FN_matrix}.
\section{Feasibility and Optimal Pool Selection for Linear and Square Array Group Testing}
\label{sec:GroupTestingMethods}

We focus on two group testing protocols: linear array testing and square array testing. These are non-adaptive methods with only two stages of testing (including a follow-up individual test) that are easy to understand and implement in practice. Because they only have two stages, the bulk of the testing can proceed in parallel, reducing the amount of time from sample collection to test result. This is important when using group testing for prevalence control, since reducing this delay reduces the number of people that an infectious person infects.

We first introduce these two methods and then compare their performances. Specifically, we compare the number of tests needed, the number of false negatives, and the number of false positives under different parameter regimes of testing capacity and prevalence rate. 

\subsection{Group Testing Protocols and Notation}
\label{subsec:GroupTestingIntro}
Suppose the total number of people to test is $N$ and the desired pool size is $n$. The linear array testing method has group size $n$, and the square array testing method has group size $n \times n$. We assume $N$ is a multiple of $n^2$ so that all samples can be arranged into linear/square arrays with pool size $n$. This assumption is valid when we consider a large population size $N$, since the fraction of individuals that cannot fit into a linear/square array in the population approaches is negligible compared to $N$.

\begin{figure}[htb]
    \centering
    \includegraphics[width=.4\textwidth]{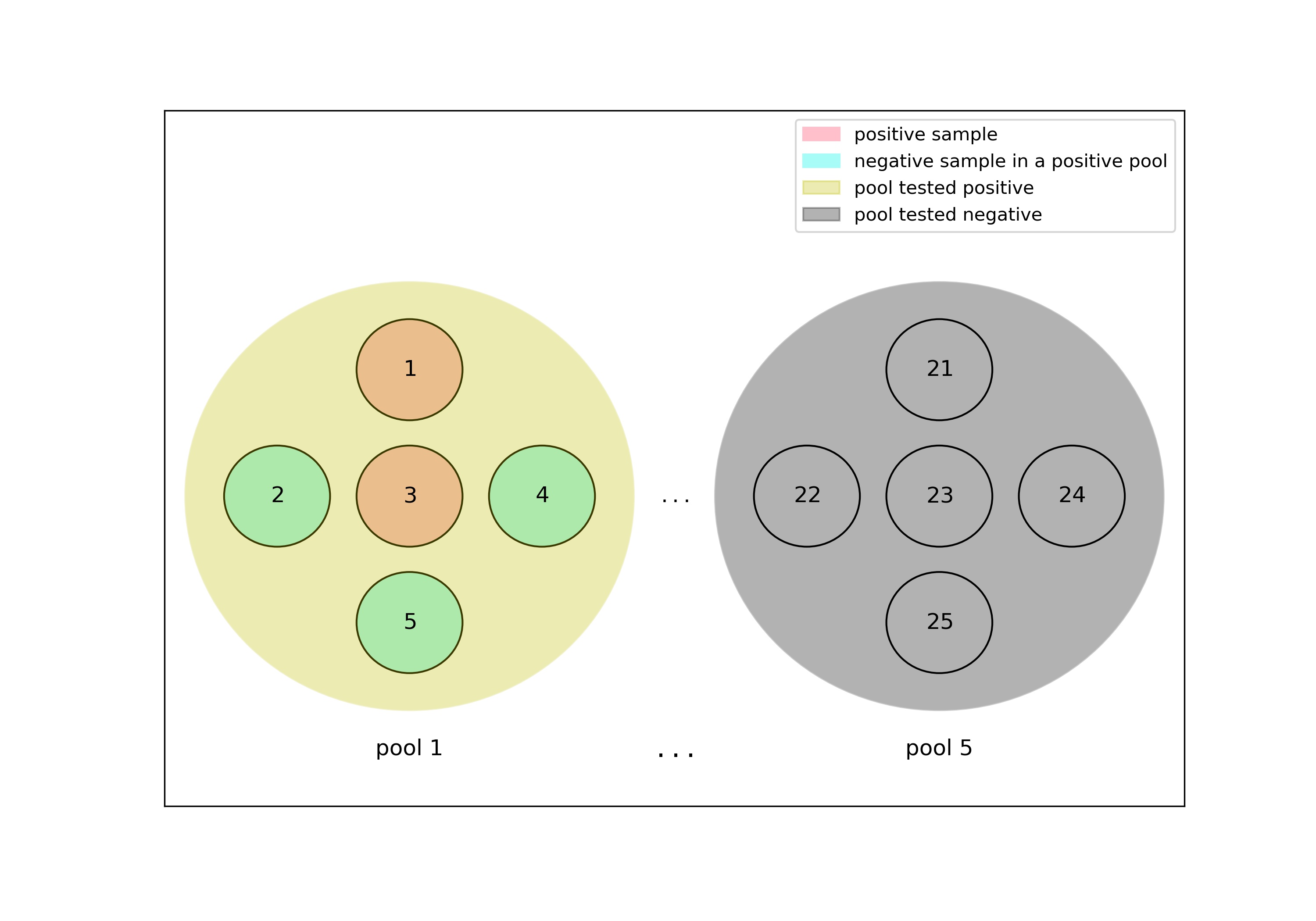}
    \caption{Placement of samples in a linear array of pool size $5$. Samples 1 and 3 are positive, causing pool 1 to test positive. Follow-up individual tests on samples in pool 1 can help identify positives 1 and 3. Samples in pools that test negative, such as those in pool 5, are deemed negative and not tested individually.}
    \label{fig: linear array}
\end{figure}

In the linear array testing, we form $ \frac{N}{n}$ linear array groups and perform pooled testing on each group. Figure \ref{fig: linear array} shows a linear array of size $n=5$. If a pool tests positive, we will test each person in the pool individually to identify the positive sample(s) within the group. Note that in the linear array testing, we split one sample into two subsamples, of which one is used for pooled testing and the other for possible follow-up individual test. 

\begin{figure}[htb]
    \centering
    \includegraphics[width=.4\textwidth]{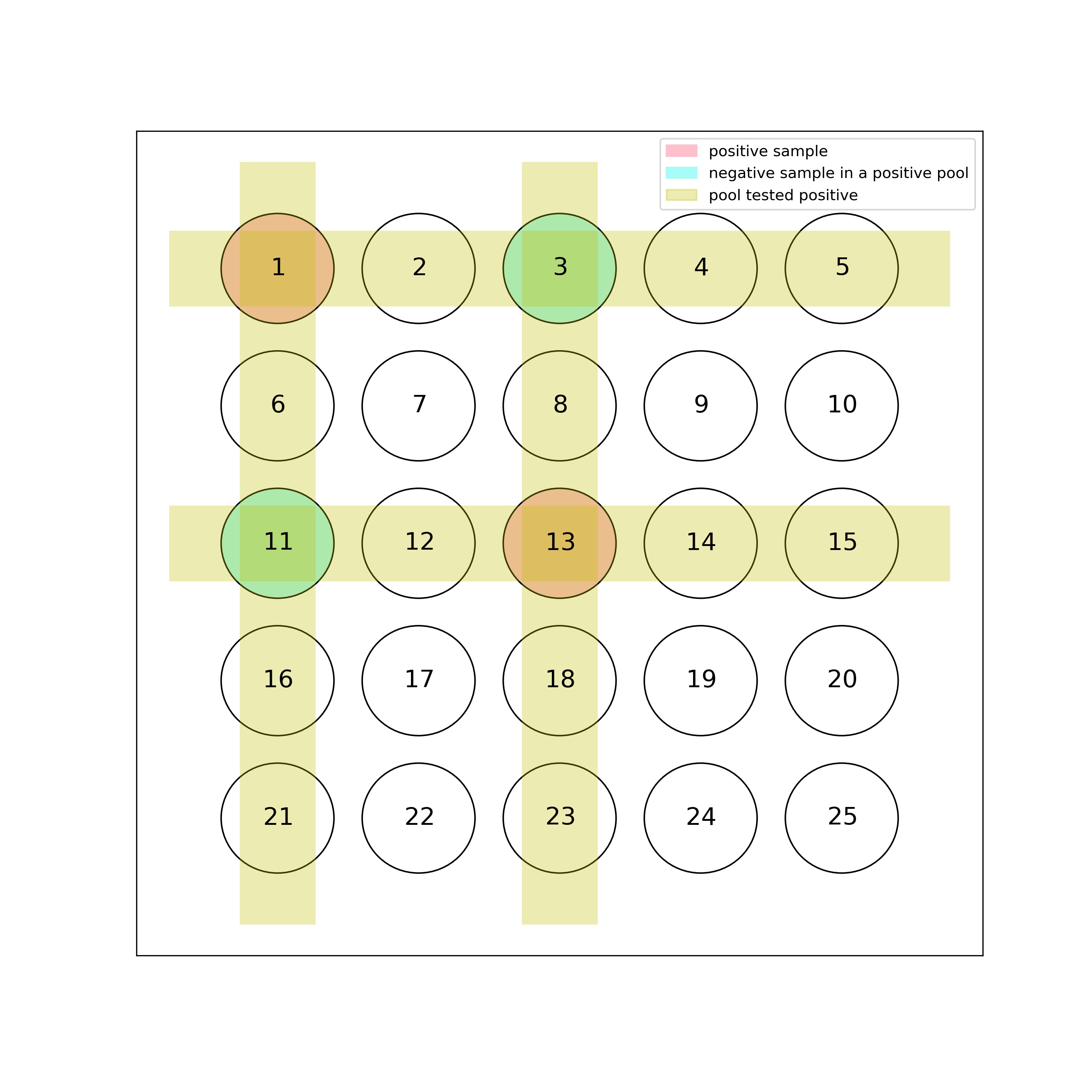}
    \caption{Placement of samples in a $5 \times 5$ square array. Samples 1 and 13 are positive. With reasonable test accuracy, it is likely that the first and the third row pools and the first and the third column pools will test positive. With follow-up individual tests on samples at the intersection of positive pools, we can declare samples 1 and 13 positive and samples 3 and 11 negative.}
    \label{fig: square array}
\end{figure}

In the square array testing, we form $\frac{N}{n^2}$ square array groups of size $n \times n$. A sample is deemed as suspicious if both its row and its column pools test positive. See Figure \ref{fig: square array} for a simple illustration. We split one sample into three subsamples, of which one is used for row testing, one is used for column testing, and the remaining one is reserved for possible follow-up individual test if the sample is deemed as suspicious.

Here we formally define the following notations:
\begin{itemize}
    \item Population size: $N$.
    \item Prevalence rate: $p$. Each individual is infected independently with probability $p$.
    \item Pool size in group testing: $n$. It is the number of samples that are pooled together in one PCR test. 
    \item Maximum pool size: $\Bar{n}$. Pooling more than $\Bar{n}$ samples together is difficult in practice (e.g. \cite{hsiang2010pcr}). We set $\bar{n}=100$.
    \item Testing capacity per person: $C$. This is the number of PCR tests available per person.
    \item Number of infected people in a pool of size $n$: $d$. Note that $d$ follows a binomial distribution with parameters $n$ and $p$.
    \item False positive rate of a PCR test: $q$. False positives in a PCR test are generally caused by false liquid handling and sample contamination. Here we assume that $q$ is a universal constant regardless of the pool size of the pooled test.
    \item False positive tolerance: $\widetilde{C}$. The population will not accept the number of false positive per person higher than this because quarantine without a good reason is very hard to bear in reality.
    \item Square array of pool size $n$: $A_{n \times n}$. $A_{i,j}$ denotes the sample/individual that belongs to the $i^{\text{th}}$ row pool and the $j^{\text{th}}$ column pool. $A_{i,:}$ denotes the $i^{\text{th}}$ row pool. $A_{:,j}$ denotes the $j^{\text{th}}$ column pool.
\end{itemize}

\subsection{Joint Distribution of Pool Results}

The linear and square array methods both involve multiple tests of the same sample: either one or two pooled tests and then a potential individual followup test.

Randomness in a sample's viral load and the fact that a sample participates in multiple tests creates complex correlations across tests.
Thus, to calculate the properties of these group testing protocols, we must understand certain joint distributions over multiple events. We define these quantities here, following a similar procedure used to define $\gamma(n,d)$ (the probability that a pooled test of size $n$ with $d$ positives has a negative result) in Section~\ref{subsec:fnrPooled}.

We define the quantities $\eta(n,d)$ for linear array tests, and $\theta(n,d_1,d_2)$ and $\psi(n, d_1,d_2)$ for square array tests. Specifically, given that a positive sample belongs to a pool of size $n$ with $d$ positives we let $\eta(n,d)$ denote the probability that it tests positive in the pooled test but tests negative in the individual test. Given that a positive sample in a square array belongs to a row pool of size $n$ with $d_1$ positives and a column pool of size $n$ with $d_2$ positives, we let $\theta(n,d_1,d_2)$ denote the probability that it tests positive in both row and column pooled tests, and $\psi(n, d_1,d_2)$ the probability that it tests positive in both row and column pooled tests but tests negative in the individual test.
We assume $1\leq d,d_1,d_2\leq n$. 
These quantities can be estimated using Monte Carlo simulation analogous to the procedure described at the end of Section \ref{subsec:fnrPooled}.

\subsection{Implications of Test Correlation}
\label{sec:correlation}
Here we call out an important consequence of our more realistic model of errors in pooled testing. 
Recall that we model a single sample as being obtained from an individual and being split into multiple subsamples for pooled and follow-up individual testing.
The viral loads for each of these sub-samples are modeled as having identical viral loads, and would indeed have extremely similar viral loads in reality.
Moreover, recall that the probability that a positive sample tests negative in a pool or in an individual test is affected largely by the viral load of the (sub-)sample in addition to idiosyncratic randomness from a given test.

As a result, positive samples with high viral loads are likely to test positive in {\it all} of the pools and individual tests they participate in. Indeed, if there were no idiosyncratic component to the probability of a pooled test testing positive and instead the test result were entirely determined by whether the viral load in the sample was above a threshold, and then we would see that a positive sample would either test positive in {\it all} pools or in {\it no pools}. (We ignore the possibility that another positive sample participated in one of these pools, which is unlikely when prevalence is small.) Thus, under our model, there is significant positive correlation in test results across pooled tests.  Moreover, because the absence of dilution effect makes individual tests more sensitive, a positive sample that tests positive in its pools is quite likely to test positive in its followup test (when prevalence is low and thus the positive pooled result is unlikely to be due to another positive sample).

As a result, the probability that a positive sample tests in all pools and its individual followup test is not that much smaller than the probability it tests positive in a single pooled test. 
This stands in sharp contrast to what one would expect from assuming errors are independent across tests: under this assumption, the probability that a sample would test positive in all tests declines exponentially in the number of tests performed.

\begin{theorem}
Consider a model identical to ours but in which the results from pooled tests and followup individual tests are all conditionally independent given which samples contain virus.  The probability that a true positive tests negative is lower under our model than in this alternative model, for both linear and square array protocols.
The probability that a true negative tests positive is unchanged.
\end{theorem}
We show this in detail in Appendix \ref{Appendix D}.

This has important consequences for adding additional tests to reduce false positives in a testing protocol. False positives are typically due to laboratory contamination from other samples in the lab. Contamination that occurs after sub-samples are taken is likely to be much less correlated across multiple pooled tests. As a result, requiring a sample to test positive in two confirmatory individual tests rather than one would significantly reduce the false positive rate. Because of the correlation due to viral load, this can be done without significantly increasing the false negative rate.

To combat contamination affecting the original sample, one could consider taking two separate samples from an individual and storing them separately. This may somewhat reduce the correlation in viral loads across pools that an individual participates in but is likely to allow even more robust control of false positives without a significant degradation in false negatives.

\subsection{Expected Number of Tests, False Negatives, and False Positives}
\label{subsec:Expressions}
We derive closed-form expressions for the expected number of tests per person, false negatives per person, and false positives per person for both linear array and square array methods. The derivations can be found in Appendices \ref{Appendix A} and \ref{Appendix B}.

With a slight abuse of notation, here we let $\gamma(n,0)$ denote the probability that a pool of size $n$ containing no positive samples tests negative. By definition, $\gamma(n,0)=1-q$. Then, the expected number of tests per person for linear array testing is
\begin{gather}
\begin{split}
M^{L}(n)  = \frac{1}{n} + \sum_{d=0}^n  \bigg(1-\gamma(n,d)\bigg )\tbinom{n}{d} (1-p)^{n-d}p^d.
\end{split}
\label{eq: ML}
\end{gather}

The expected number of false negatives per person for linear array testing is
\begin{gather}
\begin{split}
F^{L}(n) &= \sum_{d=1}^n \left(\eta(n,d)+\gamma(n,d)\right) \tbinom{n-1}{d-1} (1-p)^{n-d}p^d. 
\end{split}
\label{eq: FNL}
\end{gather}

The expected number of false positives per person for linear array testing is
\begin{gather}
\begin{split}
\widetilde{F}^{L}(n) = q(1-p)\sum_{d=0}^{n-1} (1-\gamma(n,d)) \tbinom{n-1}{d}p^d  (1-p)^{n-1-d}.
\end{split}
\label{eq: FPL}
\end{gather}

The expected number of tests per person for the square array testing is

\begin{gather}
\begin{split}
M^{S}(n) = \frac{2}{n} +p\sum_{d_1=1}^{n} \sum_{d_2=1}^{n} \theta(n,d_1,d_2)  \tbinom{n-1}{d_1-1}\tbinom{n-1}{d_2-1}\\p^{d_1+d_2-2}(1-p)^{2n-d_1-d_2} + (1-p) \bigg( \sum_{d=0}^{n-1}\\ (1-\gamma(n,d))\tbinom{n-1}{d}p^d(1-p)^{n-1-d} \bigg)^2.
\end{split}
\label{eq: MS}
\end{gather}

The expected number of false negatives per person for square array testing is
\begin{gather}
\begin{split}
F^{S}(n) =  p \bigg(1 - \sum_{d_1=1}^{n} \sum_{d_2=1}^{n} (\theta(n,d_1,d_2) -\psi(n,d_1,d_2))\\ \tbinom{n-1}{d_1-1}\tbinom{n-1}{d_2-1}p^{d_1+d_2-2}(1-p)^{2n-d_1-d_2} \bigg).
\end{split}
\label{eq: FNS}
\end{gather}

The expected number of false positives per person for square array testing is
\begin{gather}
\begin{split}
\widetilde{F}^{S}(n) = q (1-p) \bigg(\sum_{d=0}^{n-1} (1-\gamma(n,d)) \tbinom{n-1}{d}p^{d}(1-p)^{n-1-d} \bigg)^2. 
\end{split}
\label{eq: FPS}
\end{gather}

\begin{figure*}[htb]
    \centering
    \begin{subfigure}[b]{0.325\textwidth}
         \centering
         \includegraphics[width=\textwidth]{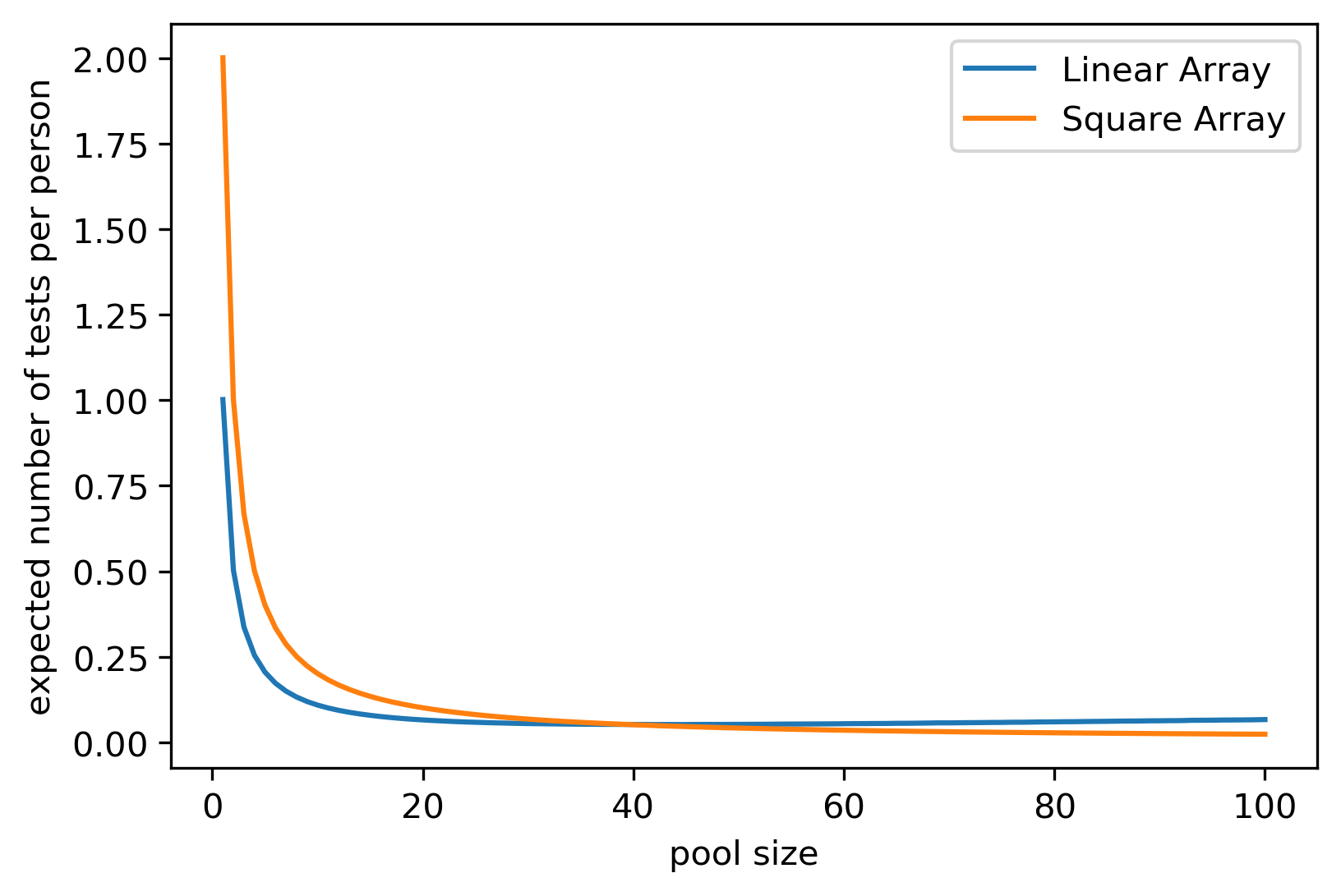}
         \caption{}
         \label{fig: comparison between two-a}
     \end{subfigure}
     \begin{subfigure}[b]{0.325\textwidth}
         \centering
         \includegraphics[width=\textwidth]{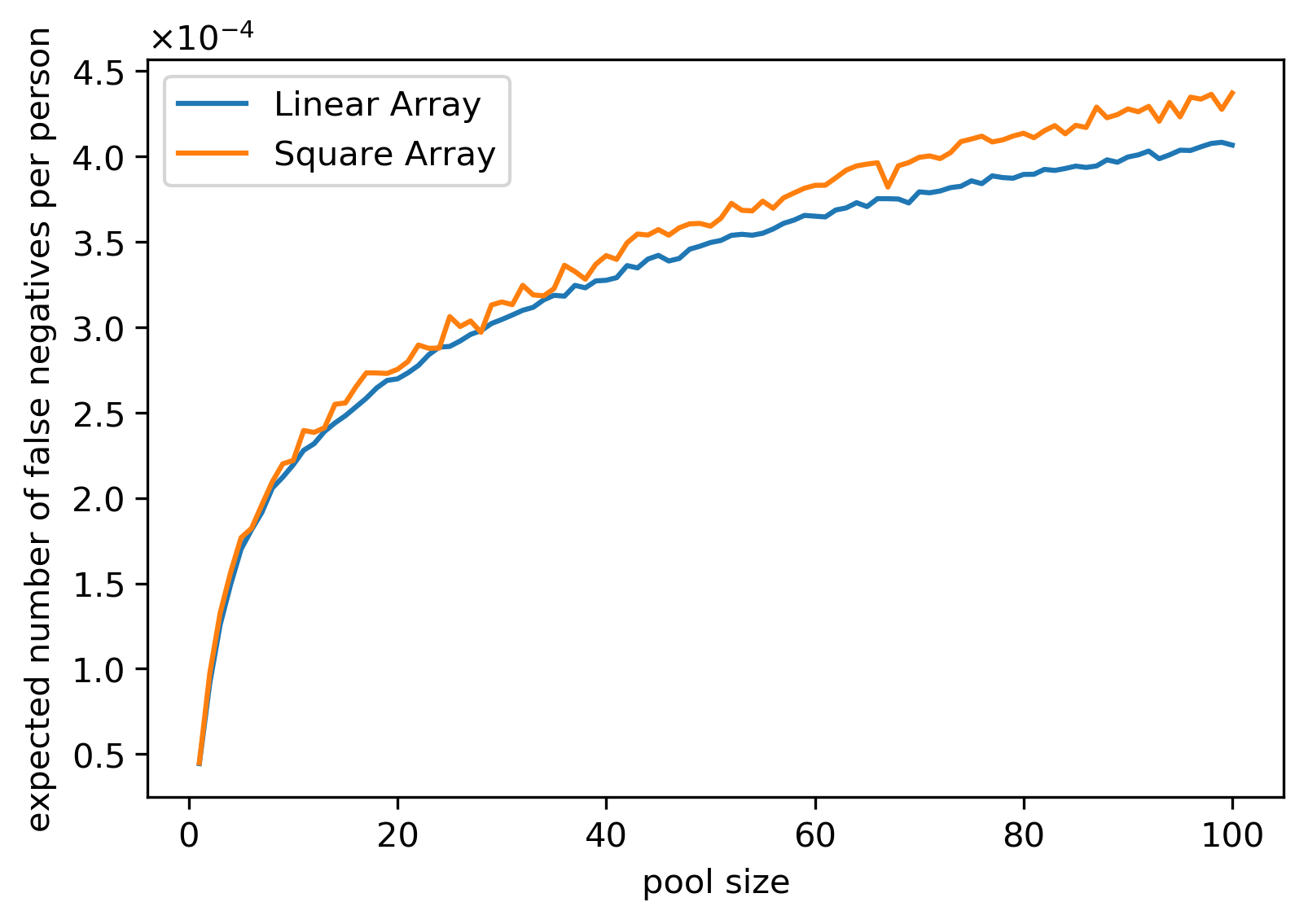}
         \caption{}
         \label{fig: comparison between two-b}
     \end{subfigure}
     \begin{subfigure}[b]{0.325\textwidth}
         \centering
         \includegraphics[width=\textwidth]{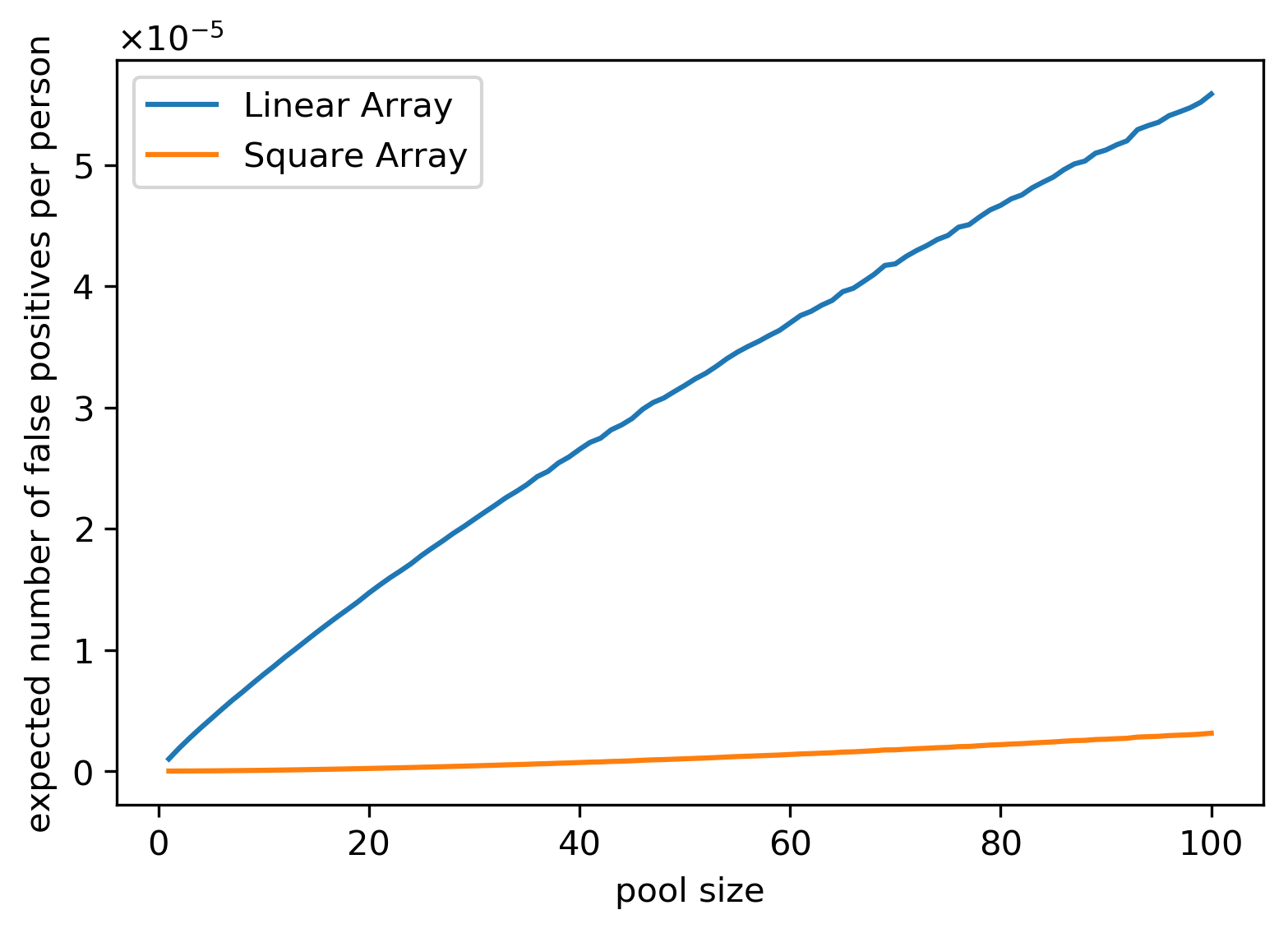}
         \caption{}
         \label{fig: comparison between two-c}
     \end{subfigure}\\
    \begin{subfigure}[b]{0.4\textwidth}
         \centering
         \includegraphics[width=\textwidth]{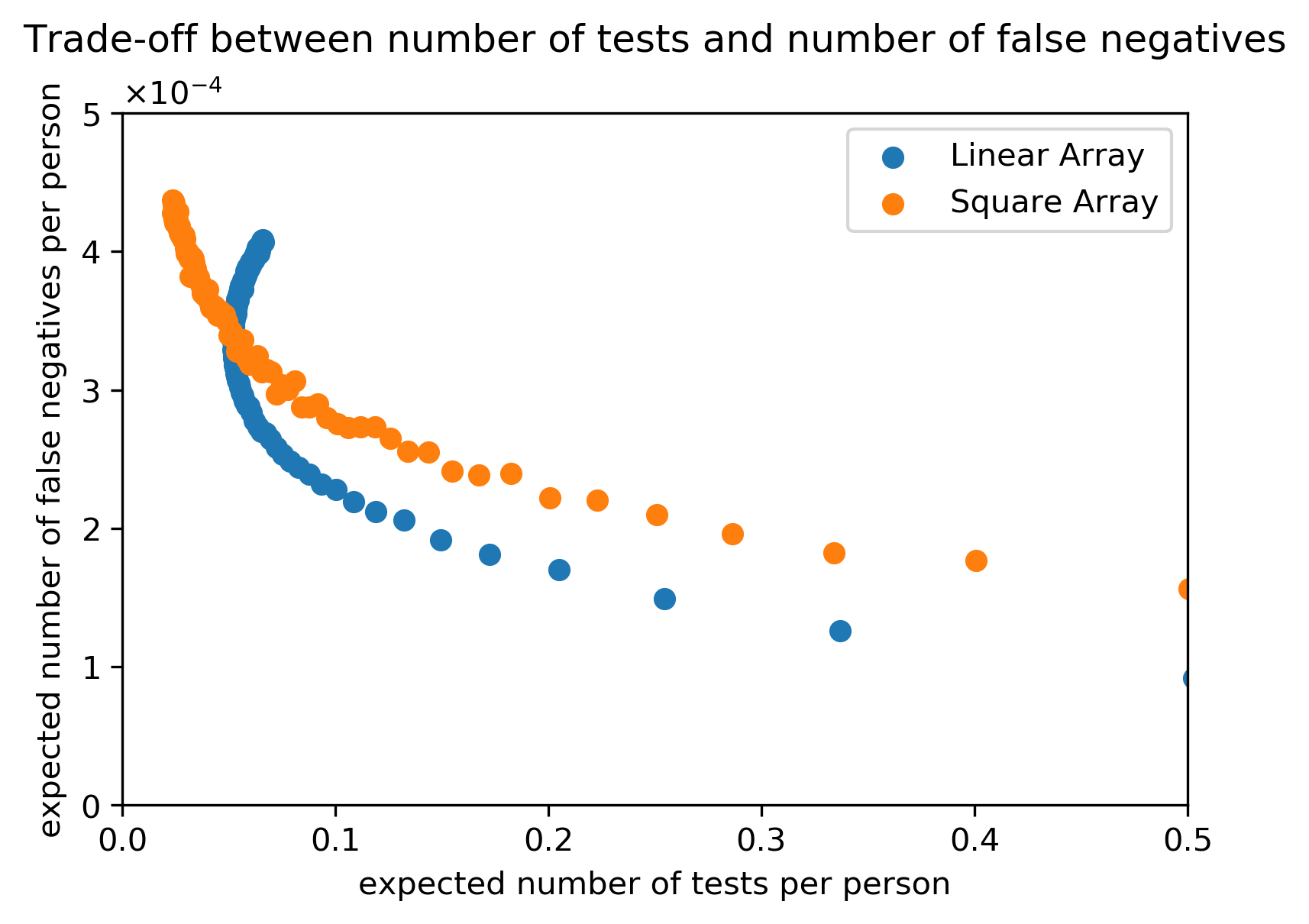}
         \caption{}
         \label{fig: comparison between two-d}
     \end{subfigure}
     \begin{subfigure}[b]{0.4\textwidth}
         \centering
         \includegraphics[width=\textwidth]{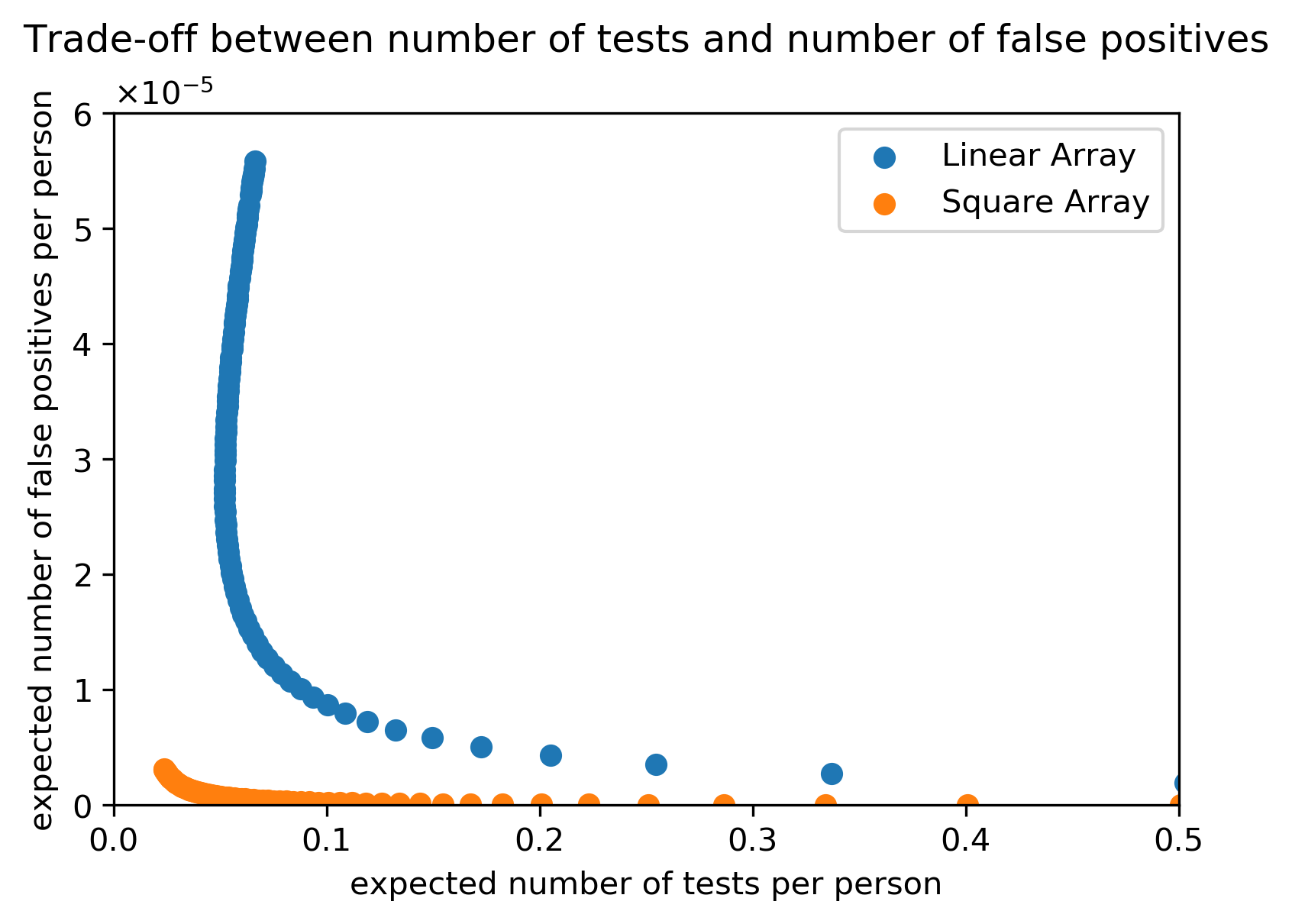}
         \caption{}
         \label{fig: comparison between two-e}
     \end{subfigure}
    \caption{Comparison between the two methods when $p=10^{-3}$ and $q=10^{-3}$. 
    In the top 3 plots (a,b,c) we show the expected number per person of tests, false negatives, and false positives as a function of the pool size for the linear and square array.
    In the bottom two plots (d) and (e), we plot the same data where each point corresponds to a pool size and a testing method (linear array or square array), and show its expected number of false negatives/positives per person versus its expected number of tests per person.}
    \label{fig: comparison between two}
\end{figure*}

Figure \ref{fig: comparison between two} (a)-(c) shows the number of tests per person, the number of false negatives per person, and the number of false positives per person with pool size $n$ ranging from 1 up to a maximum size $\bar{n}=100$ for the two methods, respectively.
It holds fixed $p=10^{-3}$ and $q=10^{-3}$. 
Figure \ref{fig: comparison between two} (d)-(e) show the same data, but varying the pool size implicitly and plotting each two error rate against the expected number of tests per person.

Foreshadowing the next section,
Figure \ref{fig: comparison between two}d-e
show that there is a trade-off between the number of tests and the error rates for both methods. 
In this example, 
the square array is able to achieve a smaller number of expected tests per person than the linear array method.
It also has lower false positive rates, due to the fact that a sample needs to test positive 3 times (twice in pools, once individually) to be declared positive under the square array method but only needs to test positive twice under the linear array.
If, however, the linear array method is able to achieve a given test capacity, then it has fewer false negatives.

\subsection{Feasibility and Optimal Pool Size}
\label{subsec:Comparison}

Here we consider the selection of pool size as an optimization problem. Varying the pool size $n$, we seek to minimize the expected number of false negatives per person tested $F^{(\cdot)}(n)$, subject to the expected number of tests per person $M^{(\cdot)}(n)$ not exceeding a given level of testing capacity per person $C$ and the expected number of false positives per person $\widetilde{F}^{(\cdot)}(n)$ not exceeding a given level of false positive tolerance $\widetilde{C}$. 

Formally, we write this as,
\begin{gather}
    \begin{aligned}
        & \underset{n \in \{1,2,\cdots,\Bar{n} \}}{\text{minimize}}
        & & F^{(\cdot)}(n) \\
        & \text{subject to}
        & & M^{(\cdot)}(n) \leq C,
        & & \widetilde{F}^{(\cdot)}(n) \leq \widetilde{C}.
    \end{aligned}
    \label{opt: section 4}
\end{gather}
where we use superscripts '$L$' for linear array testing and '$S$' for square array testing. 

These constraints reflect the fact that we face the shortage of testing kits in practice, and that we would not want to quarantine a large fraction of the population for no good reason.
In addition, by varying constraints in this optimization problem we  trace a Pareto frontier over test capacity used and the two types of errors: false positives and false negatives. In the next section, we use this Pareto frontier within the design of a larger asymptomatic screening strategy.

We say that a group test method is "feasible" if there exists a pool size $n$ such that the corresponding expected number of tests per person is no more than $C$ and the expected number of false positives per person is no more than $\widetilde{C}$, and "infeasible" otherwise. 

We solve (\ref{opt: section 4}) by first pre-eliminating infeasible pool sizes using the fact that (\ref{eq: ML}) and (\ref{eq: MS}) are lower-bounded by $\frac{1}{n}$ and $\frac{2}{n}$, respectively. Then we exploit that (\ref{eq: FNL}) and (\ref{eq: FNS}) are upper-bounded by $q(1-p)$ to see if we can drop the constraint $\widetilde{F}^{(\cdot)}(n) \leq \widetilde{C}$. Finally, we complete enumeration over the remaining pool sizes $n$ up to maximum pool size $\bar{n}=100$, evaluating the objective and constraints using the expressions 
(\ref{eq: ML})-(\ref{eq: FPS}).

Figure~\ref{fig: comparison between two} shows the 
performance of optimal pool sizes in terms of the expected number of false negatives per person while
varying the testing capacity per person $C$
and prevalence $p$, 
holding the false positives per pooled test $q=10^{-3}$ 
and constraint $\widetilde{C}=10^{-4}$ fixed.
The infeasible region for each of the two group testing methods is left blank. 
It compares the linear array, square array, and a benchmark: individual testing. 

\begin{figure*}[htb]
    \centering
    \begin{subfigure}[b]{0.325\textwidth}
         \centering
         \includegraphics[width=\textwidth]{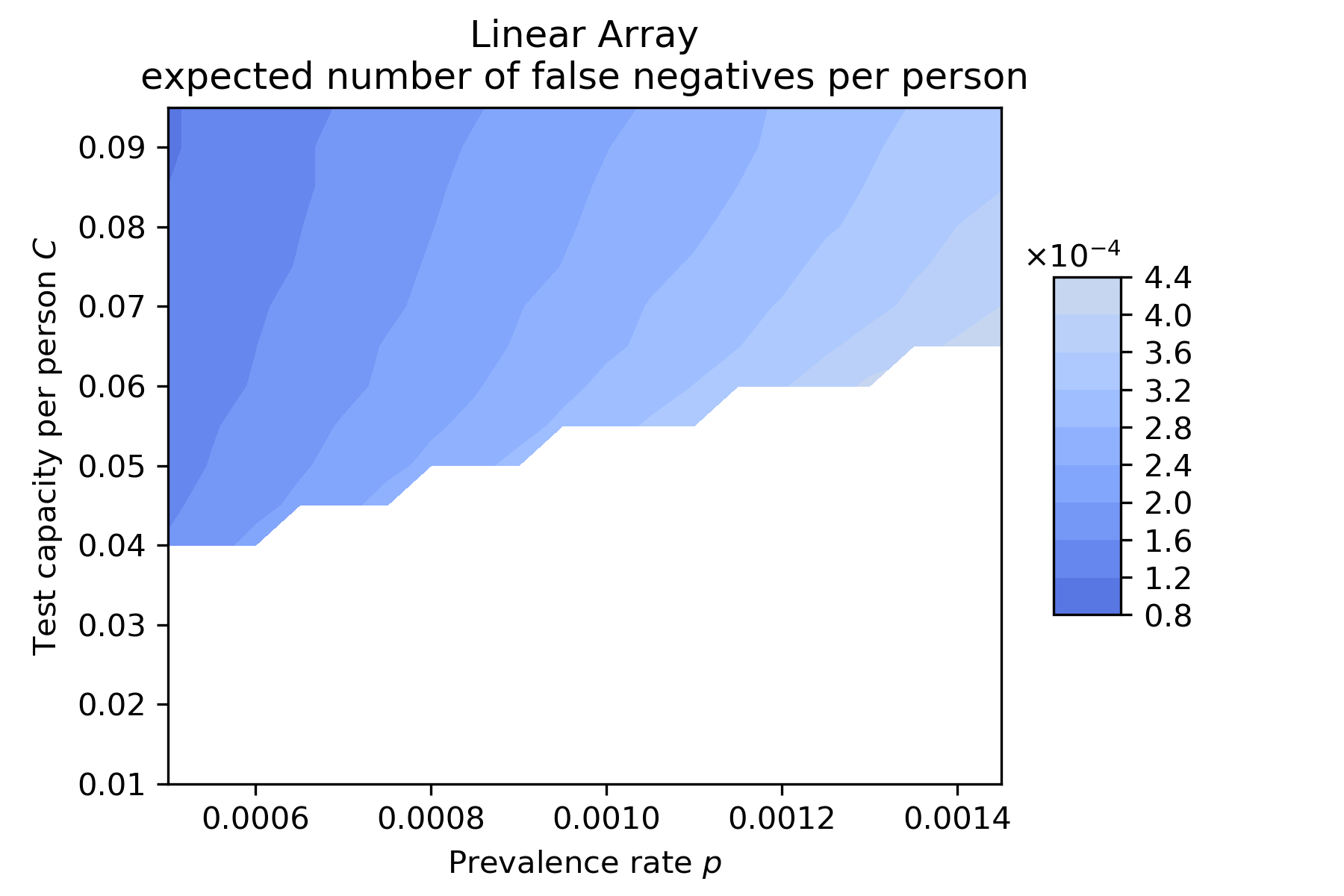}
         \caption{}
         \label{fig:method comparison-a}
     \end{subfigure}
     \begin{subfigure}[b]{0.325\textwidth}
         \centering
         \includegraphics[width=\textwidth]{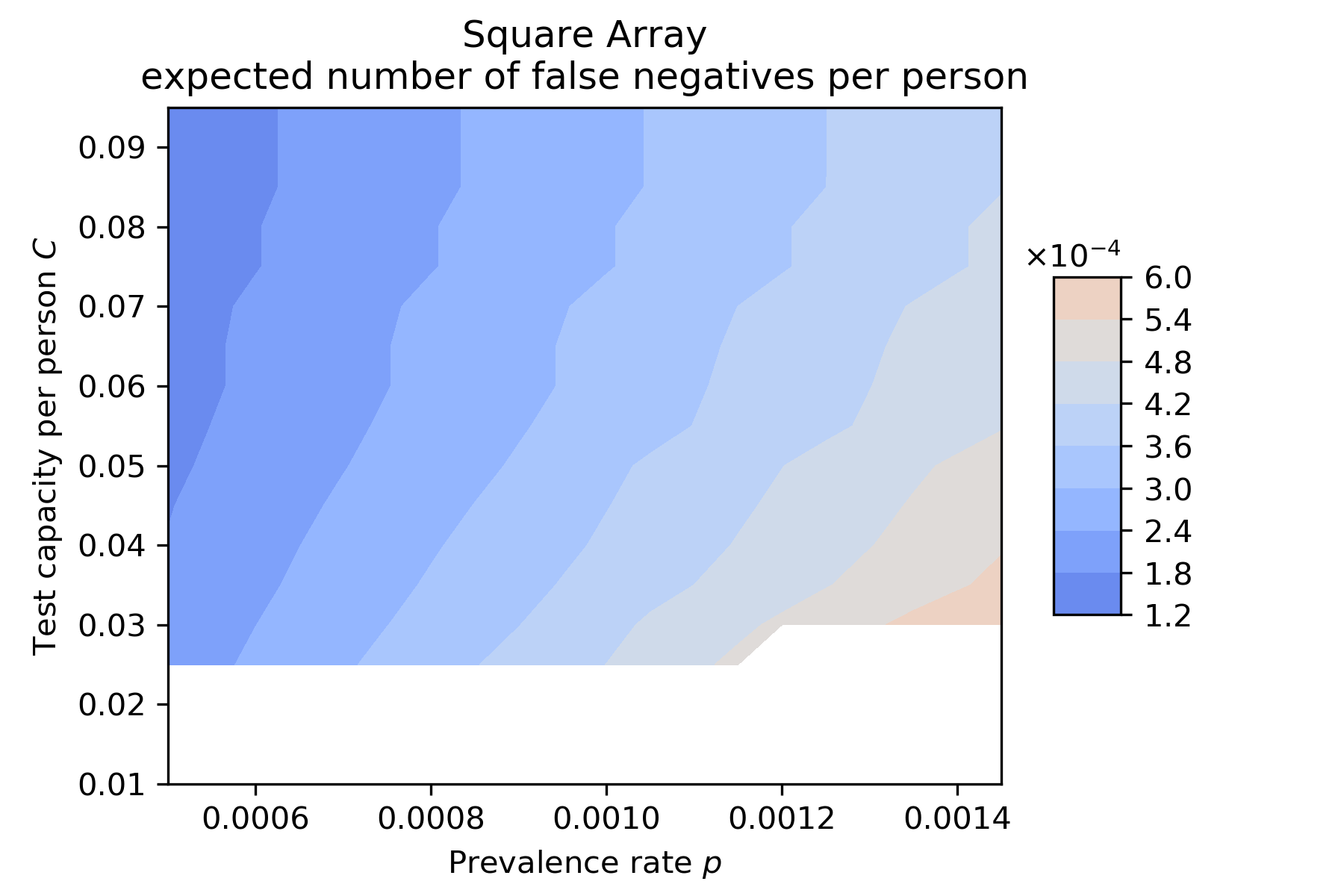}
         \caption{}
         \label{fig:method comparison-b}
     \end{subfigure}
     \begin{subfigure}[b]{0.325\textwidth}
         \centering
         \includegraphics[width=\textwidth]{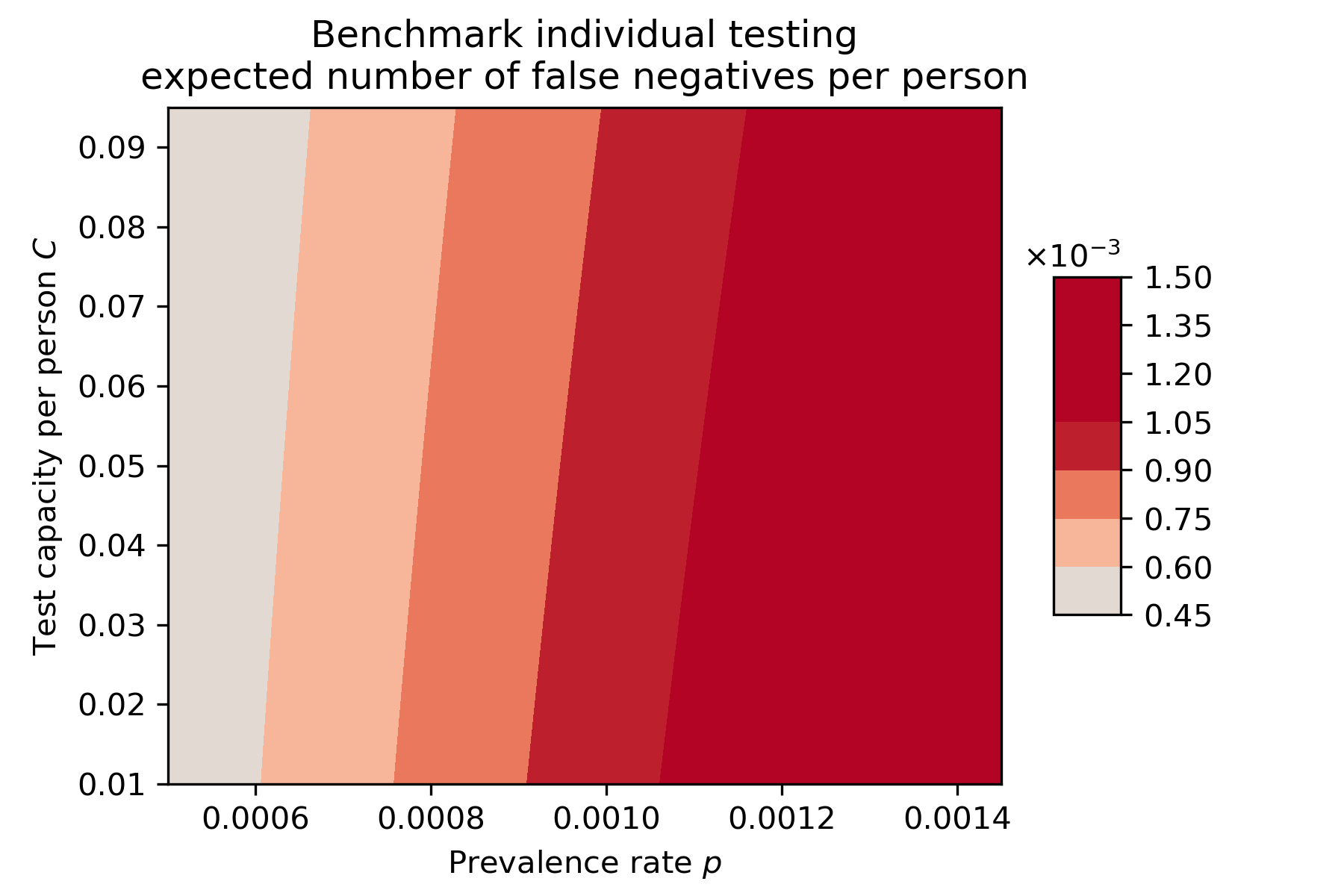}
         \caption{}
         \label{fig:method comparison-c}
     \end{subfigure}
    \caption{Comparison of the expected number of false negatives between (a) linear array group testing, (b) square array group testing, and (c) benchmark individual testing, for varying values of $p$ and $C$. Here, we use $\widetilde{C}=10^{-4}$ and $q=10^{-3}$. The infeasible region (the region in which the required test capacity cannot be achieved while testing the full population) is left blank for the two group testing methods. The individual testing benchmark satisfies the test capacity constraint but violates the false positive constraint.}
    \label{fig: method comparison}
\end{figure*}

In the individual testing benchmark, 
we randomly select a fraction $C$ of the population for which to conduct individual testing and ignore the rest of the population. 
The option to not test some individuals causes all parameters to be feasible for the test capacity constraint. This benchmark, however, violates the false positive constraint since $q>10^{-4}$.

Both group testing methods perform much better than the individual-testing-only benchmark individual. The square array has a larger feasible region than the linear array, 
confirming our earlier argument. Within the feasible region, the linear array testing achieves fewer false negatives than the square array. 

We do not focus on analysis of the impact of $\widetilde{C}$ and $q$ on the feasible region. As argued in Section~\ref{sec:correlation} and illustrated by Figure~\ref{fig: comparison between two}, the expected number of false positives per person is extremely low when $q=10^{-3}$ and $p=10^{-3}$ in all cases except the linear array with a large pool size.

\section{Repeated Asymptomatic Screening for Population-level Prevalence Control}
\label{sec:TestingCycle}

In this section, we consider testing a large population in a community such as college. Due to limited daily testing capacity, we can only test the whole population in a testing cycle of multiple days. We propose a testing-quarantine-infection model, where group testing is conducted at the beginning of each day and people who test positive will be quarantined, while the infection keeps spreading among the non-quarantined population. Due to the cost of dynamically adjusting pool size, we fix the pool size in each testing cycle, and only recompute the optimal pool size at the beginning of each cycle. 
The pool size is chosen to minimize the expected number of false negatives according to the prevalence rate at the beginning of the testing cycle. An important aspect of this model is that the number of false negatives will directly affect the number of people quarantined, which further impacts the prevalence rate over time. 

First, let us describe the primary assumptions:
\begin{itemize}
    \item On each day in a testing cycle, the group testing protocol is applied to a subset of the population which is selected uniformly at random from the untested individuals. 
    \item We use a simplified model for the infection process, where prevalence increases deterministically and exponentially without intervention.
    \item We assume the tests are conducted at the beginning of the day, and results can be revealed immediately. Following test results, people who test positive are assumed to be quarantined, either by self-quarantine or hospitalization. Thus, those who test positive will be removed from the whole population at the beginning of the day after the testing.
\end{itemize}

Let us now introduce notations for the model. The meaning of these parameters and how they relate to the model dynamics will be clarified shortly. We begin with the high-level model primitives:
\begin{itemize}
    \item $N_{total}$ is the total number of individuals in the community.
    \item $T$ is the number of time periods (days) in our time horizon.
    \item $l$ is the number of days in any one testing cycle. Note that we only consider testing cycle lengths which are shorter than the time horizon, i.e. $l\le T$.
    \item $\alpha$ is the daily growth rate of infection. 
    \item $p_0$ is the initial prevalence rate of the disease at the beginning of the time horizon.
    \item $\alpha_{out}$ is the proportion of the susceptible population that gets infected each day from interaction with the outside community. 
    \item $C^+$ is our upper bound on the expected total number of PCR tests performed in a day.  Note this is in contrast to the notation $C$ used in earlier sections, which refers to a bound on tests per person.  We use a bound on the total number of tests in this section because the population size changes over time due to individuals being quarantined.
\end{itemize}

The dynamics of each day proceed in two stages: a testing stage and an infection stage. In the testing stage, a subset of the population is tested and those who test positive will be quarantined accordingly. In the infection stage, people who are free and infectious will continue to infect the free and susceptible people in the population. 

Recall that multiple testing cycles of length $l$ occur over the time horizon of length $T$. For each day $t=1,\dots,T$ in the time horizon, let $N_t^{test}$ denote the number of individuals who are tested on that day, which is defined as follows:
\begin{gather*}
    N^{test}_{t}=\begin{cases}
        \lceil\frac{N_{t_0}}{l}\rceil, \ & t \equiv 1,2,\cdots, l-1 \mod l\\
        N_{t_0}-\lceil\frac{N_{t_0}}{l}\rceil(l-1), & t\equiv l \mod l,
        \end{cases}
\end{gather*}
where $t_0$ is the first day of the testing cycle containing day $t$, and $N_{t_0}$ is the total free population on that day.

We use the following variables to track the state of the system in each time period. Note that a variable with a bar on top refers to the not-yet-tested population, while a variable without the bar refers to the already-tested population.
\begin{itemize}
    \item $N_t$ is the total free population at the beginning of time $t$, i.e. the number of individuals who have not been quarantined, at time $t$.  
    \item On day $t$, at the beginning of the testing stage: $\bar{Z}_t$ is the total free population that have not yet been tested within the current testing cycle; $Z_t$ is the total free population that has been tested within the current testing cycle.
    \item On day $t$, at the beginning of the testing stage: $X_t$ is the total number of infected individuals in the already-tested free population; $\bar{X}_t$ is the number of infected individuals in the not-tested free population.
    \item On day $t$, at the beginning of the infection stage: $Y_t$ is the total number of infected individuals in the already-tested free population; $\bar{Y}_t$ is the number of infected individuals in the not-tested free population.
\end{itemize}

We consider testing $N_{total}$ people in $l$ days, and each day we test $N^{test}_t$ people. At the beginning of day $t$, we randomly choose $N^{test}_t$ people from those who have not been tested yet, and conduct group testing (linear array or square array methods with follow-up individual tests) under a limited testing capacity. After group testing, we mark those who already get tested, and quarantine those who test positive, and return the rest of the people back to the population. People who have been tested will not be tested again in this testing cycle. 

Let us now describe the dynamics of the testing-quarantine-infection model in more detail. The initial conditions are summarized as follows:
\begin{itemize}
    \item At the beginning of the time period, the number of positive individuals in the non-testing free population are sampled according to the specified prevalence:
    \begin{gather}
        \bar{X}_1\sim \mathrm{Binomial}(N_{total}, p_0).
    \label{eq:X1bar}
    \end{gather}
    \item On the first day $t=1$, the free non-tested population is equal to the entire population, and the tested population is equal to zero:
    \begin{gather}
        \bar{Z}_1=N_{total}, \quad Z_1=0.
    \label{eq:Z1_Z1bar}
    \end{gather}
\end{itemize}

The dynamics on each day $t$ proceed as follows. $N_t^{test}$ individuals are selected uniformly at random from the non-tested population of size $\bar{Z}_t$.  Let $I\leq \bar{X}_t$ be the total number of infected individuals in the sampled subpopulation. Let $Q_I\leq I$ be the number of infected individuals that the protocol quarantines (i.e. true positives) and let $Q_S$ be the number of susceptible individuals that the protocol quarantines (i.e. false positives). The number of positive individuals at the beginning of the infection stage for the already-tested subpopulation and the not-tested free population are updated as:
\begin{gather}
    Y_t = X_t + I - Q_I, \quad \bar{Y}_t = \bar{X}_t - I.
\label{eq:Y_Ybar}
\end{gather}

Note that the dynamics of the testing stage fundamentally depend on the quarantine decisions made by the testing protocol (as referenced via $Q_I$ and $Q_S$ in equation (\ref{eq:Y_Ybar}). Our detailed testing protocol model discussion is contained in Section \ref{sec:GroupTestingMethods}.

The infection stage occurs after the testing stage and it models the spread of the disease among the unquarantined population. We assume that daily growth rate of infection $\alpha$ is the within-community daily transmission rate, meaning that each unquarantined infected individual transmits the disease to $\alpha$ new individuals each day. Specifically, the number of new infections within the community is a binomial random variable:
\begin{gather}
    A\sim \mathrm{Binomial}(Y_t + \bar{Y}_t, \alpha),
\label{eq:A}
\end{gather}
where $Y_t + \bar{Y}_t$ is the total number of free and infected individuals.

In addition, the free and susceptible population can carry the disease from contact with the outside population. At the beginning of the infection stage, the number of free and susceptible individuals $FS$ is given by the difference between the original free population $N_t$ and the total quarantined population on this day as well as the total free and infected population on this day: 
\begin{gather}
    FS = N_t - Q_S - Q_I - Y_t - \bar{Y}_t.
\label{eq:FS}
\end{gather}

The number of new outside infections $O$ is also a binomial random variable:
\begin{gather}
    O\sim \mathrm{Binomial}(FS,\alpha_{out}).
\label{eq:O}
\end{gather}

We assume that the total number of new infections $A+O$ is spread uniformly at random across the free and susceptible population.  Let $J_t$ be the resulting number of new infections which occur in the already-tested population, and let $\bar{J}_t$ be the number of new infections which occur in the not-yet-tested population.  

If $t+1$ is in the same testing cycle as day $t$, then the infection counts $X_{t+1}$ and $\bar{X}_{t+1}$ at the start of the testing stage for the next day are then calculated as:
\begin{gather}
    X_{t+1} = Y_{t+1} + J_{t+1}, \quad \bar{X}_{t+1} = \bar{Y}_{t+1} + \bar{J}_{t+1},
\label{eq:X_Xbar_next}
\end{gather}
and the total population counts for each group are updated as:
\begin{gather}
    Z_{t+1} = Z_t + N_t^{test} - Q_I - Q_S, \quad \bar{Z}_{t+1} = \bar{Z}_t - N_t^{test}.
\label{eq:Z_Zbar_next}
\end{gather}

On the other hand, if day $t+1$ is the beginning of a new testing cycle, then the state variables reset since there are no individuals in the already-tested group. We then have:
\begin{gather}
    Z_{t+1}=X_{t+1}=0, \quad \bar{Z}_{t+1} = Z_{t} + \bar{Z}_t - Q_S - Q_I,
\label{eq:Z_reset}
\end{gather}
and 
\begin{gather}
    \bar{X}_{t+1} = \bar{Y}_{t} + Y_t + J_t + \bar{J}_t.
\label{eq:X_reset}
\end{gather}

We also run simulations to determine the optimal cycle length $l^*$ from the outset of the dynamics. This optimization also proceeds by simulating the dynamics for each feasible cycle length $l$ and choosing the cycle length with the minimum final prevalence rate. The algorithm for choosing the optimal cycle length is summarized in Algorithm \ref{algorithm1}.

\begin{algorithm}
\SetAlgoLined
\SetKwInOut{Input}{input}\SetKwInOut{Output}{output}
\Input{$N_{total}$, $p_0$, $\alpha$, $C^+$, $T$}
\Output{optimal testing cycle length $l^{*}$}
\For{$l \leftarrow 1$ \KwTo $T$}{
\For{replication $\leftarrow 1$ \KwTo $100$}{
Simulate the population-level infection dynamics described in Section \ref{sec:TestingCycle} through a sequence of equations (\ref{eq:X1bar}) to (\ref{eq:X_reset}) using a cycle length $l$ and parameters $N_{total}$, $p_0$, $\alpha$, $C^+$\;
Record final prevalence at the end of the time horizon\;
}
Compute the average final prevalence rate at the end of the given time period\;
}
Return the optimal testing cycle length $l^{*}$ that yields the smallest average final prevalence rate\;

\caption{Optimizing cycle length in the testing-quarantine-infection model.}
\label{algorithm1}
\end{algorithm}

Finally, let us elaborate on details of the testing protocol we use in our testing-quarantine-infection model. 

In our simulations, we use the same pool size in our group testing protocol for each day in the same testing cycle, but we re-optimize for a new pool size at the beginning of each testing cycle. That is, we re-solve the optimal pool size optimization problem (\ref{opt: section 4}) at the beginning of each testing cycle using information about the current prevalence rate and the cycle length.

Further, we use a slight modification of the model described in Section \ref{sec:GroupTestingMethods}, where the model allows us to gracefully handle test population sizes which do not perfectly fit into a square array of the desired size. Suppose we have a total of $N_t^{test}$ samples to test on each day. Since $N_t^{test}$ is large, we choose the pool size using the optimization (\ref{opt: section 4}) for the square array method.

In doing so, we consider using a mixed method of linear array and square array. More specifically, we fill $X$ samples in $Y$ complete square arrays, and fill the remaining $Z$ samples row by row in a square of the same size. The square arrays are tested using the square array method, and each row of the incomplete square is tested using the linear array method. This lets us improve our testing efficiency and make the tests more convenient to implement without making changes to the lab setting for the square array tests. Moreover, since we consider large population, the outcomes are insensitive to our choice for the method for testing the leftover. 
% We conduct $\lfloor\frac{N_t^{test}}{n^2}\rfloor$ square array tests and for the remaining $N_t^{test}-n^2\lfloor\frac{N_t^{test}}{n^2}\rfloor$ samples, we will do $\left\lfloor\frac{N_t^{test}-n^2\lfloor\frac{N_t^{test}}{n^2}\rfloor}{n}\right\rfloor$ linear array tests of pool size $n$ and a linear array test of pool size $N_t^{test}-n\lfloor\frac{N_t^{test}}{n}\rfloor$ if $N_t^{test}-n\lfloor\frac{N_t^{test}}{n}\rfloor\neq0$.
The details are shown in Appendix \ref{Appendix C}.
As expected, because $N$ is large, we found that the results were almost identical.

Note that the optimization formulation (\ref{opt: section 4}) we use to recompute the optimal pool size at the beginning of each testing cycle is formulated with respect to a standard square array protocol, rather than the mixed square array and linear array protocol that is implemented in our simulations.  In the appendix we consider an exact optimization formulation for the mixed array protocol (\ref{opt: section 5}).  In practice, we observe that both formulations produce nearly identical pool size selections.

Details of the mixed array testing protocol within one testing cycle are summarized in Algorithm \ref{algorithm2}. Assume that this cycle begins at day $t+1$.

\begin{algorithm}[htb]
\SetAlgoLined
Input: Number of people to be tested $N_{t}$, estimate of initial prevalence rate $p_t$, testing capacity $C^+$, cycle length $l$\;
%Run the testing-quarantine-infection model (i.e., Algorithm \ref{algorithm1})\;
Output: %optimal testing cycle length $l^{*}$, optimal pool size $n_{t}^{*},t=1,\cdots,l^{*}$\;
decisions for each people get tested in this testing cycle \;
Solve optimization problem (\ref{opt: section 4}) based on $N_t, p_t, C^+$ and obtain the optimal pool size $n^*$ for the entire testing cycle\;
Set $i=t+1$\;
\While{$i \leq t+l$}{
    In day $i$, take swab from each individual\;
    Form $\lfloor \frac{N^{test}_i}{ (n^{*})^2} \rfloor$ square array of size $n^{*} \times n^{*}$. Pools of size $n^{*}$ are created from each row and each column\;
    Conduct RT-qPCR test for each pool\; 
    Conduct individual test for samples at the intersection of positive row and positive column\;
    Conduct linear array test for the remaining $N^{test}_i-\lfloor \frac{N^{test}_i}{(n^{*})^2} \rfloor  (n^{*})^2$ people. All except the last linear array are using size $n^*$, the size of the last array is depend on the number of remaining people\;
    Make decisions on each individuals according to the test results.
    %Quarantine people who test positive, and return people who test negative back to the community\;
    Set $i=i+1$.
}
\caption{Testing protocol within one testing cycle.}
\label{algorithm2}
\end{algorithm}

\section{Case Study and Sensitivity Analysis}
\label{sec:CaseStudy}
In this section, we use the testing protocol we present in Section \ref{sec:TestingCycle} to contain the spread of the disease in a community. 

This case study aims to provide guidance to decision makers about how to use group testing protocols to perform population-level screening with limited testing capacity.

The formal details of our population-level infection dynamics model are specified in Section \ref{sec:TestingCycle}.
Our model has inputs which make assumptions about the size of the community, the initial prevalence rate in the community, and the rate at which the epidemic grows.  Table \ref{Table 1} states the value we assign to these input variables, separated into optimistic, nominal, and pessimistic scenarios. 

The testing cycle length $l$ varies from 1 to 7, and we consider a $14$-day time horizon, which guarantees multiple (at least $2$) complete cycles can be observed. We record the prevalence rate after each day and we use $100$ replications for simulating each feasible testing cycle length in each scenario. 

\begin{table*}[htbp]
    \centering
        \caption{Parameter settings for three scenarios: optimistic case, nominal case, and pessimistic case.}
    \begin{tabular}{|c|c|c|c|}
    \hline
      Scenario Name & Optimistic case & Nominal case & Pessimistic case \\
      \hline
       $N_{total}$: Population size  & 100000& 100000&100000 \\
       $T$: Time Horizon  & 14 &14& 14\\
       $p_0$: Initial prevalence rate & 0.001& 0.005&0.01 \\
       $C^+$: The total number of testing capacity & 3000 & 3000 & 3000\\
       $q$: The false positive rate of each PCR test & 0.01 & 0.01 & 0.01\\
       %Within-community contacts per day & 1 & 5 & 10 \\
       %Transmissions per contact & 0.02 & 0.02 & 0.02 \\
       $\alpha$: Daily transmissions per free infected & 0.1&0.2 &0.3 \\
       $\alpha_{out}$: Daily outside infection rate & 0.00002& 0.0005&0.003 \\
       \hline
    \end{tabular}
    \label{Table 1}
\end{table*}

\subsection{Simulation Results and Analysis}
\label{sec:Sim}
\begin{figure*}[htb]
    \centering
    \begin{subfigure}[b]{0.325\textwidth}
         \centering
         \includegraphics[width=\textwidth]{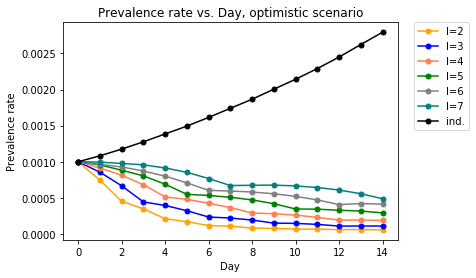}
         \caption{}
         \label{fig: prevalence-a}
     \end{subfigure}
     \begin{subfigure}[b]{0.325\textwidth}
         \centering
         \includegraphics[width=\textwidth]{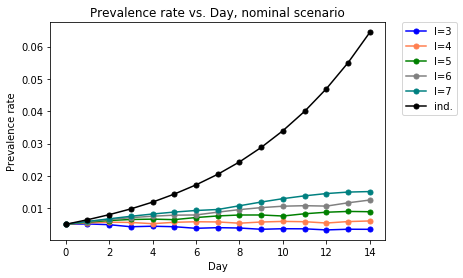}
         \caption{}
         \label{fig: prevalence-b}
     \end{subfigure}
     \begin{subfigure}[b]{0.325\textwidth}
         \centering
         \includegraphics[width=\textwidth]{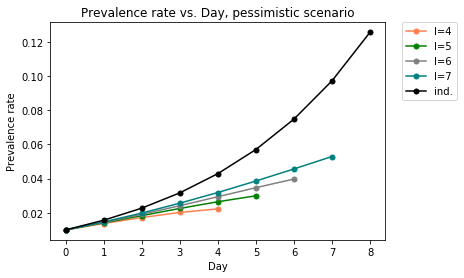}
         \caption{}
         \label{fig: prevalence-c}
     \end{subfigure}
    \caption{Prevalence rates of the mixed array method with different testing cycle length $l$ under (a) optimistic scenario; (b) nominal scenario; (c) pessimistic scenario.}
    \label{fig: prevalence}
\end{figure*}

\begin{figure*}[htb]
    \centering
    \begin{subfigure}[b]{0.325\textwidth}
    \includegraphics[width = \textwidth]{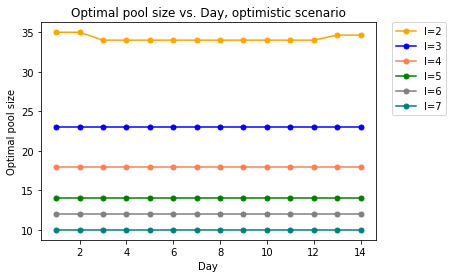}
    \caption{}
    \label{fig: optimal pool size-a}
    \end{subfigure}
    \begin{subfigure}[b]{0.325\textwidth}
    \includegraphics[width =\textwidth]{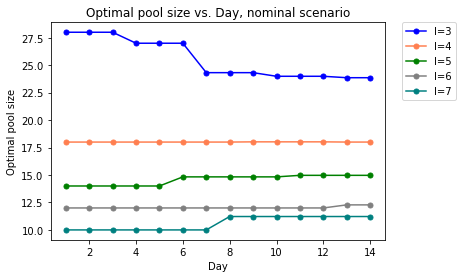}
    \caption{}
    \label{fig: optimal pool size-b}
    \end{subfigure}
    \begin{subfigure}[b]{0.325\textwidth}
    \includegraphics[width = \textwidth]{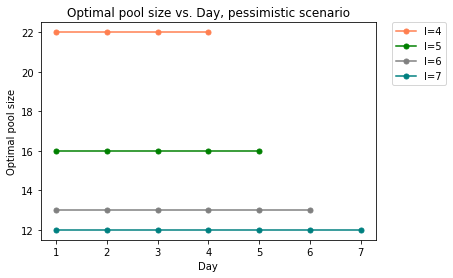}
    \caption{}
    \label{fig: optimal pool size-c}
    \end{subfigure}
    \\
    \begin{subfigure}[b]{0.325\textwidth}
    \includegraphics[width = \textwidth]{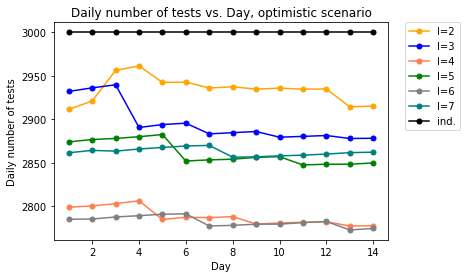}
    \caption{}
    \label{fig: optimal pool size-d}
    \end{subfigure}
    \begin{subfigure}[b]{0.325\textwidth}
    \includegraphics[width = \textwidth]{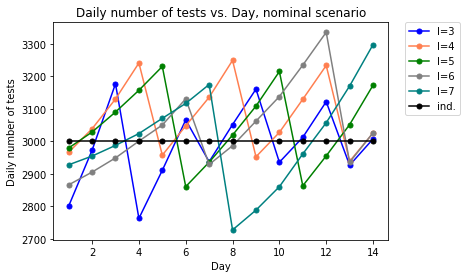}
    \caption{}
    \label{fig: optimal pool size-e}
    \end{subfigure}
    \begin{subfigure}[b]{0.325\textwidth}
    \includegraphics[width = \textwidth]{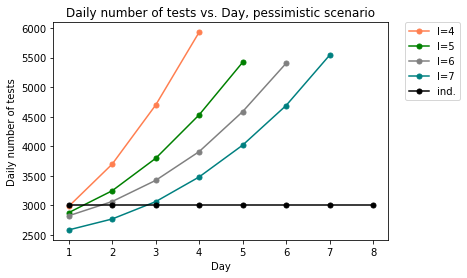}
    \caption{}
    \label{fig: optimal pool size-f}
    \end{subfigure}
    \caption{Optimal pool size of the mixed array method with different testing cycle length under (a) optimistic scenario; (b) nominal scenario; (c) pessimistic scenario; Daily number of tests of the mixed array method with different testing cycle length under (d) optimistic scenario; (e) nominal scenario; (f) pessimistic scenario;}
    \label{fig: optimal pool size}
\end{figure*}

In each scenario, we keep track of the prevalence rate each day within the $14$-day time horizon. Figure \ref{fig: prevalence} shows the prevalence rate under different cycle lengths $l$ and also the prevalence rate under individual testing. For the optimistic scenario, the cycle length $l=1$ is not feasible due to the daily testing capacity, so it is not included in Figure \ref{fig: prevalence}(a). For the nominal scenario, neither $l=1$ nor $l=2$ is feasible at the beginning, so they are not shown in Figure \ref{fig: prevalence}(b). For the pessimistic scenario, $l=1,2,3$ are not feasible at the beginning, and $l=i$ is not feasible at day $i$, $i=4,5,6,7$. Therefore, Figure \ref{fig: prevalence}(c) has a truncated pattern. The estimated final prevalence rates, total number of quarantined individuals and total number of testing are listed in Table \ref{Table 2}. The optimal pool size and the number of tests each day are shown in Figure \ref{fig: optimal pool size}.

\begin{table*}[htb]
    \caption{The estimated final prevalence rate, total number of quarantined, and total number of tests for the mixed array group testing with different cycle length $l$ and the benchmark individual testing. For pessimistic scenario, the results are based on only one testing cycle.}
    \centering
    \begin{filecontents*}{table/data2.csv}
,Testing cycle length,1,2,3,4,5,6,7,individual
Optimistic ,Final prevalence rate,NAN,5.62E-05,1.10E-04,1.87E-04,2.92E-04,4.14E-04,4.88E-04,2.79E-03
,Total quarantined,NAN,143.24,152.01,159.52,164.37,164.59,172.97,499.45
,Total test times,NAN,41076.9,40358.1,39034.41,40057.26,38954.06,40074.96,42000
Nominal ,Final prevalence rate,NAN,NAN,3.46E-03,6.05E-03,8.90E-03,1.25E-02,1.51E-02,6.44E-02
,Total quarantined,NAN,NAN,1693.53,1820.97,1955.04,1974.68,2099.43,1444.52
,Total test times,NAN,NAN,41844.36,43075.21,42668.99,42551,42119.53,42000
Pessimistic ,Final prevalence rate,NAN,NAN,NAN,2.23E-02,2.99E-02,3.98E-02,5.29E-02,5.40E-01
,Total quarantined,NAN,NAN,NAN,1467.93,1786.47,2211.33,2694.66,6129.11
,Total test times,NAN,NAN,NAN,17303.84,19869.96,23208.97,26145.92,42000
\end{filecontents*}
\csvautobooktabular{table/data2.csv}
    \label{Table 2}
\end{table*}

Figure \ref{fig: prevalence} shows that for all three scenarios, the mixed array method leads to a much lower final prevalence rate than the individual testing. However, in the pessimistic setting, the virus spreading is out of control. Hence we focus on the first two scenarios in the following. The final prevalence rate turns out to be monotonically decreasing as testing cycle decreases. As a result, the optimal testing cycle length is $l=2$ for the optimistic scenario, and the optimal testing cycle length is $l=3$ for the nominal case. In contrast, for both optimistic and nominal scenarios the prevalence rate gets out of control when individual test is used. 

\begin{figure*}[htb]
    \centering
    \begin{subfigure}[b]{0.325\textwidth}
    \includegraphics[width = \textwidth]{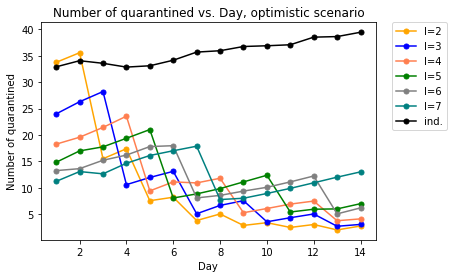}
    \caption{}
    \label{fig: number of quarantined-a}
    \end{subfigure}
    \begin{subfigure}[b]{0.325\textwidth}
    \includegraphics[width = \textwidth]{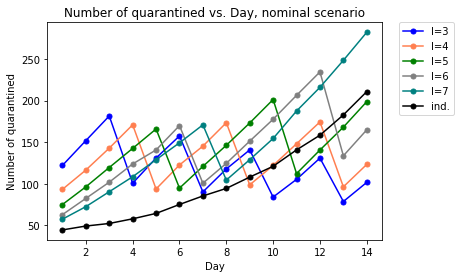}
    \caption{}
    \label{fig: number of quarantind-b}
    \end{subfigure}
    \begin{subfigure}[b]{0.325\textwidth}
    \includegraphics[width = \textwidth]{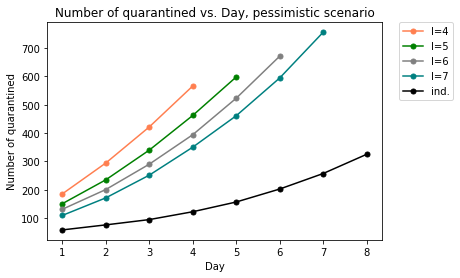}
    \caption{}
    \label{fig: number of quarantined-c}
    \end{subfigure}
    \caption{The number of quarantined people of mixed array method with different testing cycle length under (a) optimistic scenario; (b) nominal scenario; (c) pessimistic scenario}
    \label{fig: number of quarantined}
\end{figure*}

Figure \ref{fig: optimal pool size} suggests that the optimal pool size is around $10$ to $30$, which is not too large. However, considering the pool size limit $100$ we set, this is relatively small. This is because we are minimizing false negative rate in (\ref{opt: section 4}), and false negative rate becomes low as the pool size decreases. Compared with the individual testing, group testing uses much less number of tests each day. Table \ref{Table 2} shows that the number of tests is minimized when $l=4$ under the optimistic scenario. Also, from Figure \ref{fig: number of quarantined}, the number of true positives detected of individual testing is much lower than group testing in all scenarios, and correspondingly, the false positive rate of individual is higher. 

\begin{figure*}[htb]
    \centering
    \begin{subfigure}[b]{0.325\textwidth}
    \includegraphics[width = \textwidth]{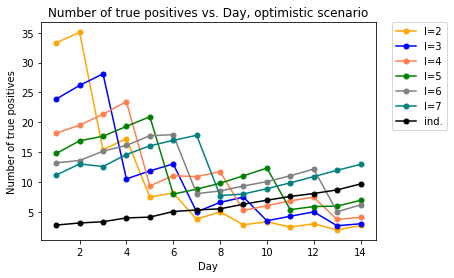}
    \caption{}
    \label{fig: number of false positives-a}
    \end{subfigure}
    \begin{subfigure}[b]{0.325\textwidth}
    \includegraphics[width = \textwidth]{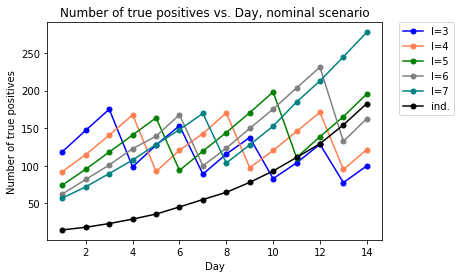}
    \caption{}
    \label{fig: number of false positives-b}
    \end{subfigure}
    \begin{subfigure}[b]{0.325\textwidth}
    \includegraphics[width = \textwidth]{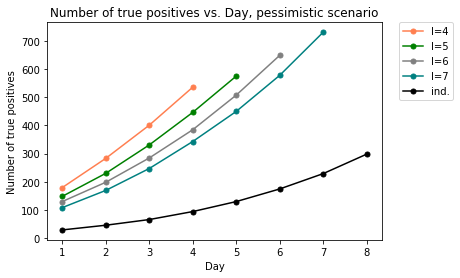}
    \caption{}
    \label{fig: number of false positives-c}
    \end{subfigure}
    \\
    \begin{subfigure}[b]{0.325\textwidth}
    \includegraphics[width = \textwidth]{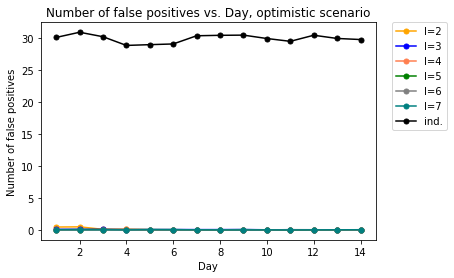}
    \caption{}
    \label{fig: number of false positives-d}
    \end{subfigure}
    \begin{subfigure}[b]{0.325\textwidth}
    \includegraphics[width = \textwidth]{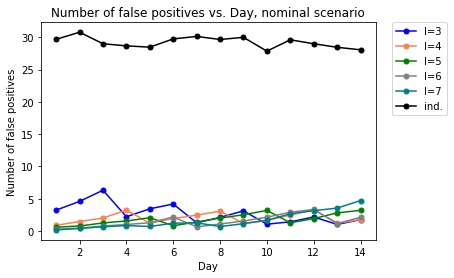}
    \caption{}
    \label{fig: number of false positives-e}
    \end{subfigure}
    \begin{subfigure}[b]{0.325\textwidth}
    \includegraphics[width = \textwidth]{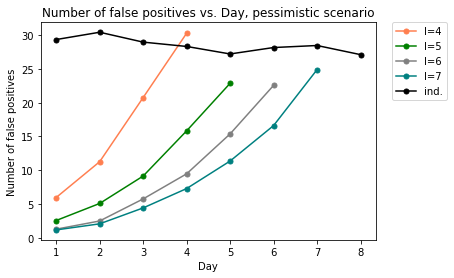}
    \caption{}
    \label{fig: number of false positives-f}
    \end{subfigure}
    \caption{Number of true positives of mixed array method with different testing cycle length under (a) optimistic scenario; (b) nominal scenario; (c) pessimistic scenario; Number of false positives of mixed array method with different testing cycle length under (d) optimistic scenario; (e) nominal scenario; (f) pessimistic scenario.}
    \label{fig: number of false positives}
\end{figure*}

\begin{figure}[htb]
    \centering
    \includegraphics[width=.4\textwidth]{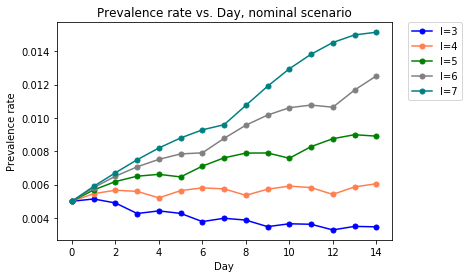}
    \caption{Detailed look for the prevalence rate curve of $l=3,4,5,6,7$ under nominal scenario.}
    \label{fig: prevalence curve}
\end{figure}

\begin{figure}[htb]
    \centering
    \includegraphics[width=.42\textwidth]{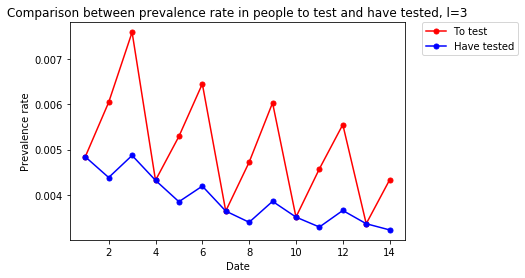}\\
    \includegraphics[width=.42\textwidth]{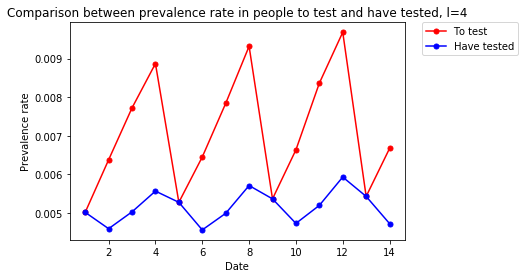}
    \caption{Comparison of prevalence rate among two subsets of all non-quarantined individuals.}
    \label{fig: prevalence comparison}
\end{figure}

In addition, it is worth-noting that the prevalence rate exhibits a periodic trend. As shown in Figure \ref{fig: prevalence curve}, if we have a detailed look of two prevalence rate curves of $l=3,4$ under nominal scenario, we can see the periodic pattern clearly. To explain this special pattern, before the testing stage of each day $t$, we consider the prevalence rate among two subsets of all non-quarantined individuals: those who has been tested and those who are going to be tested, and we plot these two prevalence rates in Figure \ref{fig: prevalence comparison}. 
Within one testing cycle, Figure \ref{fig: prevalence comparison} shows that the the prevalence rate in the about-to-test individuals becomes much higher than that in the individuals that have been tested. This will result in the increase of daily number of tests in a testing cycle, as Figure \ref{fig: optimal pool size}(e) shows, and more tests will lead to more quarantined, as shown in Figure \ref{fig: number of quarantined}, thus low prevalence rate in return. 
Therefore, with this chain reaction, we observe the prevalence rate go through a process of first increasing and then decreasing. It turns out that with every feasible testing cycle length and under every scenario, this pattern exists.

\begin{figure*}[htb]
    \centering
    \begin{subfigure}[b]{0.325\textwidth}
    \includegraphics[width=\textwidth]{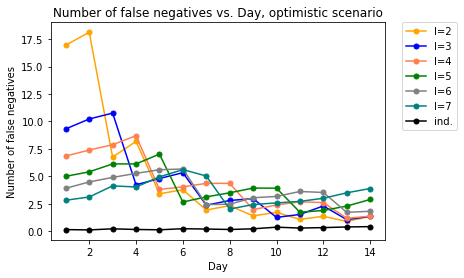}
    \caption{}
    \label{fig: number of false negatives scenario-a}
    \end{subfigure}
    \begin{subfigure}[b]{0.325\textwidth}
    \includegraphics[width=\textwidth]{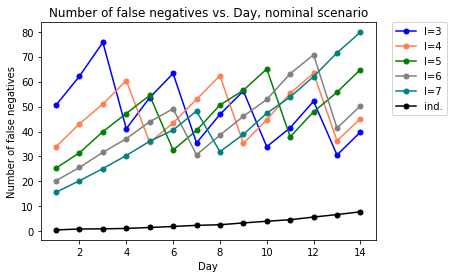}
    \caption{}
    \label{fig: number of false negatives scenario-b}
    \end{subfigure}
    \begin{subfigure}[b]{0.325\textwidth}
    \includegraphics[width=\textwidth]{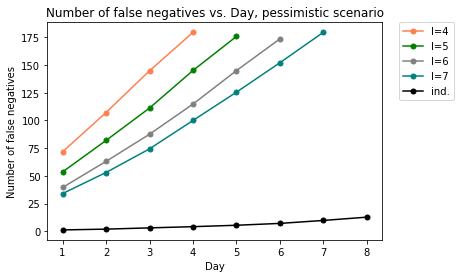}
    \caption{}
    \label{fig: number of false negatives scenario-c}
    \end{subfigure}
    \caption{Number of false negatives of mixed array method with different testing cycle length under (a) optimistic scenario; (b) nominal scenario; (c) pessimistic scenario.}
    \label{fig: number of false negatives scenario}
\end{figure*}

Furthermore, it is interesting to observe that for all three scenarios, the smaller cycle length initially has the largest false negative rate because of the large pool size, but the false negative rate drops quickly over time and becomes comparable with other cycle lengths, as shown in Figure \ref{fig: number of false negatives scenario}. Under the current parameter setting, it turns out that when $l$ is small, the larger tested population offsets the disadvantage brought by higher false negative rate, leading to the result that more infected individuals get quarantined and hence less virus carrier in the population.

\subsection{Sensitivity Analysis}
\label{subsec:Sensitivity}
In this section we vary the testing capacity parameter  $C^+$ and the false positive rate parameter  $q$, which are two key parameters for the performance of virus control other than $p_0$ and $\alpha$, which have already been considered in the three scenarios. 

\subsubsection{Sensitivity Analysis of Testing Capacity}
\label{subsubsec:Capacity}
We do sensitivity analysis on the testing capacity $C^+$ for both the nominal and pessimistic scenario.  Specifically, for the nominal scenario we run simulations changing the testing capacity value to each of the following values: $C^+= 5000, 10000, 20000$.  For the pessimistic scenario, we simulate the difference between $C^+=10000$ and $C^+=20000$.

\begin{figure*}[htbp]
    \centering
    \begin{subfigure}[b]{0.37\textwidth}
    \includegraphics[width = \textwidth]{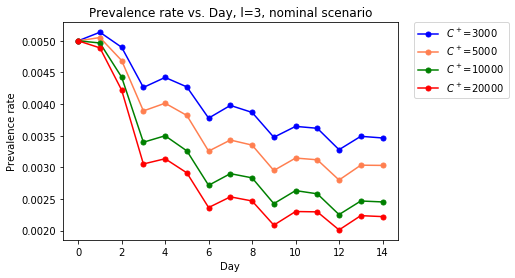}
    \caption{}
    \label{fig: prevalence rate scenario-a}
    \end{subfigure}
    \begin{subfigure}[b]{0.37\textwidth}
    \includegraphics[width = \textwidth]{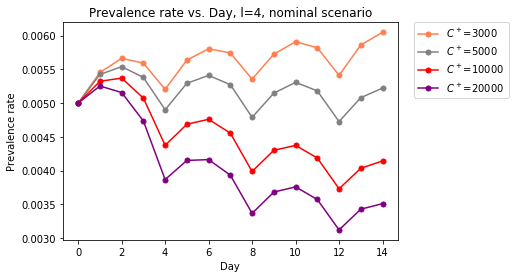}
    \caption{}
    \label{fig: prevalence rate scenario-b}
    \end{subfigure}
    \\
    \begin{subfigure}[b]{0.37\textwidth}
    \includegraphics[width = \textwidth]{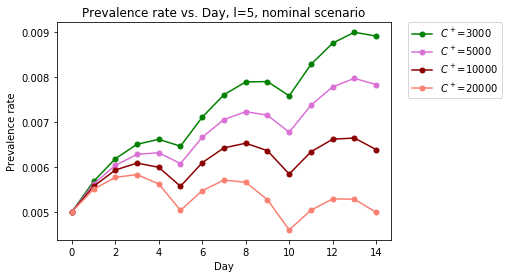}
    \caption{}
    \label{fig: prevalence rate scenario-c}
    \end{subfigure}
    \begin{subfigure}[b]{0.37\textwidth}
    \includegraphics[width = \textwidth]{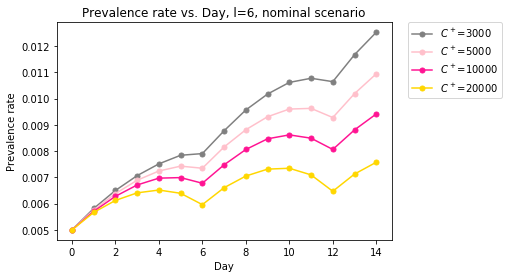}
    \caption{}
    \label{fig: prevalence rate scenario-d}
    \end{subfigure}
    \caption{Prevalence rate of different values of $C^+$ with (a) $l=3$, (b) $l=4$, (c) $l=5$, (d) $l=6$ under nominal scenario.}
    \label{fig: prevalence rate scenario}
\end{figure*}

\begin{figure*}[htb]
    \centering
    \begin{subfigure}[b]{0.37\textwidth}
    \includegraphics[width = \textwidth]{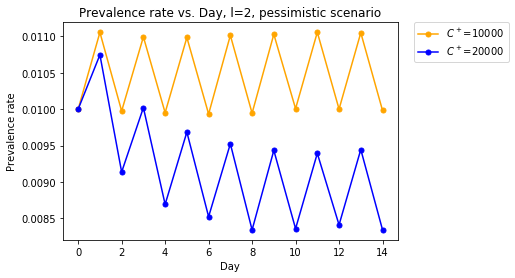}
    \caption{}
    \label{fig: prevalence rate pessimistic-a}
    \end{subfigure}
    \begin{subfigure}[b]{0.37\textwidth}
    \includegraphics[width = \textwidth]{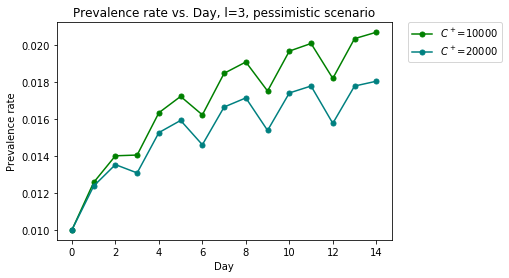}
    \caption{}
    \label{fig: prevalence rate pessimistic-b}
    \end{subfigure}
    \\
    \begin{subfigure}[b]{0.37\textwidth}
    \includegraphics[width = \textwidth]{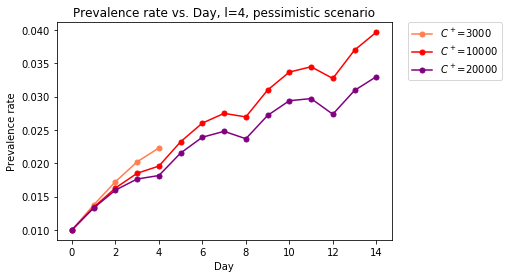}
    \caption{}
    \label{fig: prevalence rate pessimistic-c}
    \end{subfigure}
    \begin{subfigure}[b]{0.37\textwidth}
    \includegraphics[width = \textwidth]{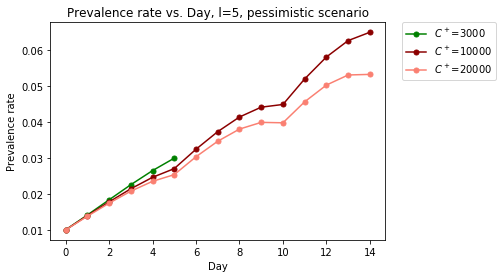}
    \caption{}
    \label{fig: prevalence rate pessimistic-d}
    \end{subfigure}
    \caption{Prevalence rate for different values of $C^+$ with (a) $l=2$, (b) $l=3$, (c) $l=4$, (d) $l=5$ under pessimistic scenario}
    \label{fig: prevalence rate pessimistic}
\end{figure*}

Figure \ref{fig: prevalence rate scenario} shows how the prevalence rate varies over time in the nominal scenario for each of the testing capacity values $C^+$, and for multiple values of the testing cycle length $l$. Figure \ref{fig: prevalence rate pessimistic} similarly shows how prevalence rate varies over time for different testing capacities $C^+$ and different testing cycle lengths $l$.

\begin{figure*}[htb]
    \centering
    \begin{subfigure}[b]{0.37\textwidth}
    \includegraphics[width = \textwidth]{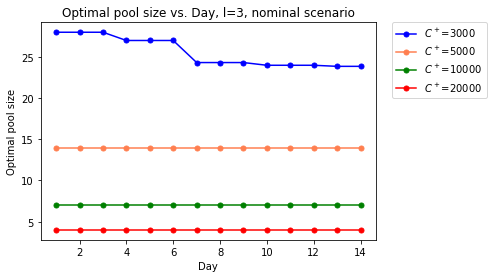}
    \caption{}
    \label{fig: everything scenario-a}
    \end{subfigure}
    \begin{subfigure}[b]{0.37\textwidth}
    \includegraphics[width = \textwidth]{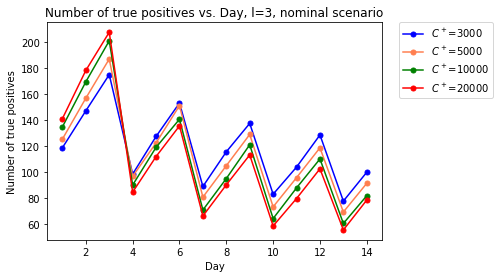}
    \caption{}
    \label{fig: everything scenario-b}
    \end{subfigure}
    \\
    \begin{subfigure}[b]{0.37\textwidth}
    \includegraphics[width = \textwidth]{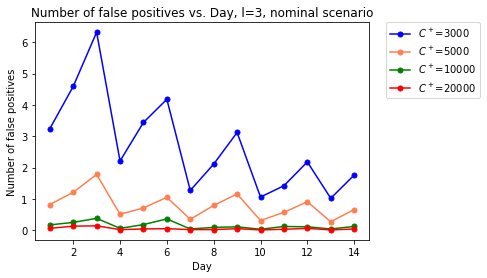}
    \caption{}
    \label{fig: everything scenario-c}
    \end{subfigure}
    \begin{subfigure}[b]{0.37\textwidth}
    \includegraphics[width = \textwidth]{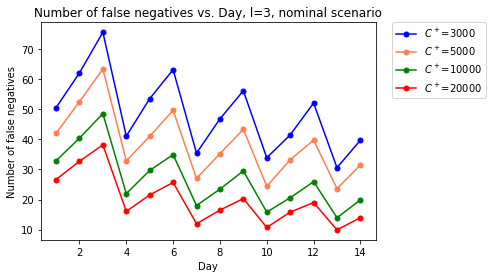}
    \caption{}
    \label{fig: everything scenario-d}
    \end{subfigure}
    \caption{(a) Optimal pool size, (b) number of true positives, (c) number of false positives and (d) number of false negatives for different values of $C^+$ under nominal scenario, $l=3$.}
    \label{fig: everything scenario}
\end{figure*}

From Figure \ref{fig: prevalence rate scenario}, we observe that changes in $C^+$ have a large effect throughout the whole time horizon. For example, in the nominal scenario simulations for a short testing cycle length $l=3$ the final prevalence rate under $C^+=3000$ is $0.0035$, whereas the final prevalence under $C^+=10000$ is $0.0025$. This is a reduction in prevalence by almost 30\%.  A similar trend can be observed for all the testing cycle length values $l$ we include.

We observe a similar effect for the pessimistic scenario, although the magnitude of the effect is less pronounced.  From Figure \ref{fig: prevalence rate pessimistic} we see that with a short testing cycle $l=2$ that final prevalence under $C^+=10000$ is $0.01$, while final prevalence under $C^+=20000$ is $0.0085$; a $15$\% reduction in prevalence.

From this simulation analysis we can draw the conclusion that increased testing capacity is a valuable resource for controlling prevalence in the population.

\subsubsection{Sensitivity Analysis of False Positive Rate}
\label{subsubsec:FPR}
\begin{figure*}[htb]
    \centering
    \begin{subfigure}[b]{0.37\textwidth}
    \includegraphics[width = \textwidth]{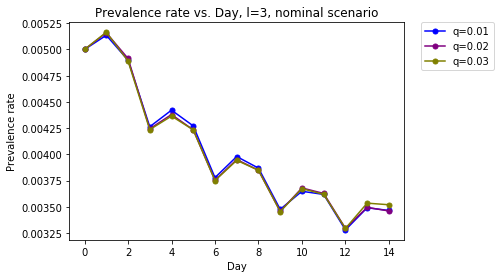}
    \caption{}
    \label{fig: prevalence rate false positive-a}
    \end{subfigure}
    \begin{subfigure}[b]{0.37\textwidth}
    \includegraphics[width = \textwidth]{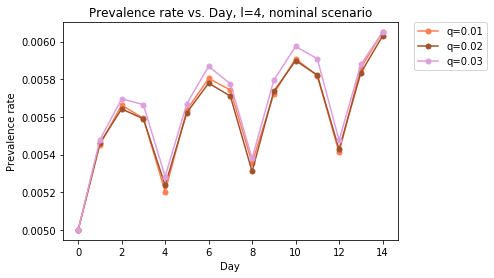}
    \caption{}
    \label{fig: prevalence rate false positive-b}
    \end{subfigure}
    \\
    \begin{subfigure}[b]{0.37\textwidth}
    \includegraphics[width = \textwidth]{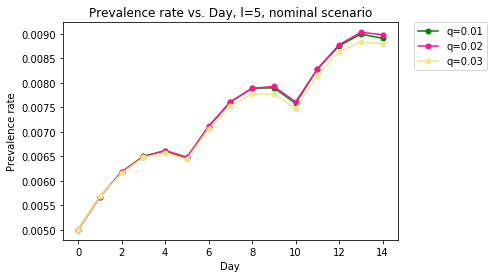}
    \caption{}
    \label{fig: prevalence rate false positive-c}
    \end{subfigure}
    \begin{subfigure}[b]{0.37\textwidth}
    \includegraphics[width = \textwidth]{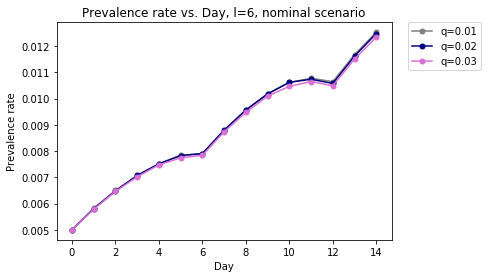}
    \caption{}
    \label{fig: prevalence rate false positive-d}
    \end{subfigure}
    \caption{Prevalence rate for different values of $q$ with (a) $l=3$, (b) $l=4$, (c) $l=5$, (d) $l=6$ under nominal scenario.}
    \label{fig: prevalence rate false positive}
\end{figure*}

\begin{figure*}[htb]
    \centering
    \begin{subfigure}[b]{0.37\textwidth}
    \centering
    \includegraphics[width=\textwidth]{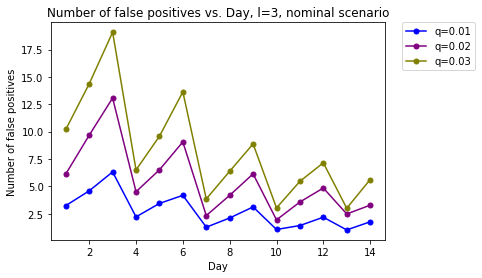}
    \caption{}
    \label{fig: false positive nominal-a}
    \end{subfigure}
    \begin{subfigure}[b]{0.37\textwidth}
    \centering
    \includegraphics[width=\textwidth]{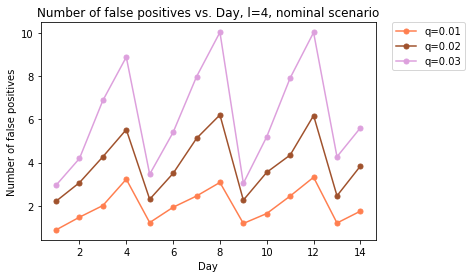}
    \caption{}
    \label{fig: false positive nominal-b}
    \end{subfigure}
    \\
    \begin{subfigure}[b]{0.37\textwidth}
    \centering
    \includegraphics[width=\textwidth]{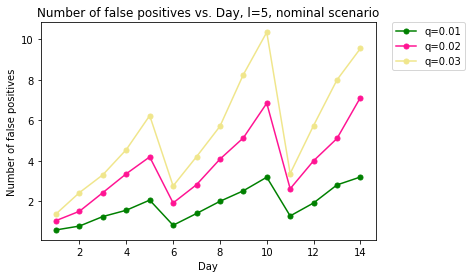}
    \caption{}
    \label{fig: false positive nominal-c}
    \end{subfigure}
    \begin{subfigure}[b]{0.37\textwidth}
    \centering
    \includegraphics[width=\textwidth]{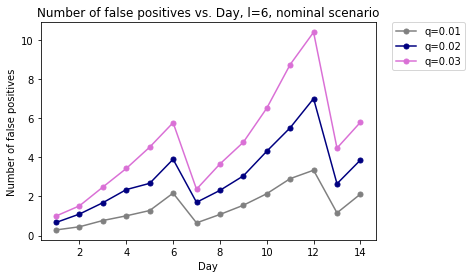}
    \caption{}
    \label{fig: false positive nominal-d}
    \end{subfigure}
    \caption{Number of false positives (i.e. uninfected individuals who are quarantined) for different values of $q$ with (a) $l=3$, (b) $l=4$, (c) $l=5$, (d) $l=6$ under nominal scenario.}
    \label{fig: false positive nominal}
\end{figure*}

Our sensitivity analysis for the false positive rate $q$ focuses
on the nominal scenario, and runs simulations setting over the range of values: $q=0.01, 0.02, 0.03$.  Our metric of interest for this study is both the prevalence rate and the daily number of uninfected individuals who enter quarantine.  Results for the prevalence rate are plotted in Figure \ref{fig: prevalence rate false positive} and results for the number of false positives are plotted in Figure \ref{fig: false positive nominal}.

From Figure \ref{fig: false positive nominal}, we can observe that the number of false positives will increase as $q$ increases, which is to be expected. Moreover, in Figure \ref{fig: prevalence rate false positive} we observe that the prevalence rate is in fact increasing for each of the testing cycle lengths $l=4,5,6$. From this we conclude that a rapid testing cycle length can have a strong effect on containing the virus spread in a close community, and that this effect dominates the effect of the false positive rate $q$ over the range of parameters we consider.

\section*{Declarations}
\subsection*{Funding}
The research was supported by the National Science Foundation under Grant CAREER CMMI-1453934 and the Air Force Office of Scientific Research under Grant FA9550-19-1-0283. 
\subsection*{Conflicts of interest}
The authors declare that they have no conflict of interest.
\subsection*{Availability of data and material}
Not Applicable.
\subsection*{Code availability}
All codes that are used in this study can be found in \href{https://github.com/Wanogon/COVID-19-Project}{https://github.com/Wanogon/COVID-19-Project}.

\appendix
\addcontentsline{toc}{section}{Appendices}
\section*{Appendices}

\section{Effect of Correlation between Tests on False Negatives}
\label{Appendix A}
Above we argued that 
a consequence of our model is that test results for infected people are correlated across the pools it participates in (including pools of size 1).
We then argued that this reduces false negatives compared with a model that assumes test results for such samples are independent. 
Here in this section, we demonstrate this formally.

\newtheorem{thm:old}{Theorem}
\begin{thm:old}
Consider a model identical to ours but in which the results from pooled tests and followup individual tests are all conditionally independent given which samples contain virus.  The probability that a true positive tests negative is lower under our model than in this alternative model, for both linear and square array protocols.
The probability that a true negative tests positive is unchanged.
\end{thm:old}
\begin{proof}
We first focus on false negatives.
Consider a true positive sample. 
Let $V$ be its viral load. 
Given a pool of size $n$ in which this (true) positive sample of interest participates, and that the positive sample has a viral load of $V$, let $r(n,V)$ be the conditional probability that this pool tests positive.  Calculating this quantity requires marginalizing across the all other $n-1$ samples in the pool, whether they are positive, their viral loads if they are, and the probability $q$ that the pool tests positive even if the viral load is not sufficient to make it do so alone.
In our model, $r(n,V)$ is increasing in $V$.

Let $Y_j$ be the event that the $j$th pool our sample of interest participates in tests positive. Let $Z$ be the event that its follow-up individual test tests positive (if this test is not actually conducted, $Z$ is the event that it {\it would} have tested positive, if it were conducted). Thus, $Y_j$ has probability $r(n,V)$ and $Z$ has probability $r(1,V)$.

Consider the linear array method. Under this method, the sample tests positive if both $Y_1$ and $Z$ occur. This occurs with probability 
\begin{equation*}
P(Y_1, Z) = E[P(Y_1,Z|V)]=E[P(Y_1|V) P(Z|V)]
\end{equation*}
since $Y_1$ and $Z$ are conditionally independent given $V$.
\cite{schmidt2014} shows that the expectation of a product of increasing functions of a random variable is larger than the product of the expectations. Since 
$P(Y_1|V)=r(n,V)$ and $P(Z|V)=r(1,V)$ are both increasing in $V$, we have 
\begin{align*}
P(Y_1, Z)
\ge E[P(Y_1|V)] E[P(Z|V)]
= P(Y_1) P(Z)
\end{align*}
which is the probability that the sample would test positive if test errors were independent across tests.

The proof for the square array method is similar, except it must also leverage the fact that results from a positive sample's two pools are conditionally independent given its viral load because no other samples that participate are in both row and column pools.

The square array identifies the sample as positive if events $Y_1$, $Y_2$ and $Z$ occur.
\begin{align*}
P(Y_1, Y_2, Z) 
&= E[P(Y_1,Y_2,Z|V)]
=E[P(Y_1|V) P(Y_2|V) P(Z|V)]\\
&\ge E[P(Y_1|V)] E[P(Y_2|V)] E[P(Z|V)]\\
&= P(Y_1) P(Y_2) P(Z)
\end{align*}
where we have used the above-described conditional independence of $Y_1$ and $Y_2$ given $V$,
\cite{schmidt2014}
and the fact that 
$P(Y_1|V) = P(Y_2|V) = r(n,V)$ and
$P(Z|V) = r(1,V)$ are both increasing in $V$.

We now turn to false positives.
As before consider a fixed sample but now suppose it is a true negative and so has $V=0$ viral load.

The probability that it tests positive under the linear array is 
$P(Y_1,Z|V=0) = P(Y_1|V=0) P(Z|V=0)$
because the pooled test result and the individual followup result are conditionally independent given the viral load.
This is the same probability that would be provided under the alternate model assuming independent errors.
The key difference between this analysis of a true negative and the above analysis of a true positive is that the viral load is deterministic given that the sample is a true negative but is random given that it is a true positive. 

The square array case is similar, and we again must use that the other samples participating in the sample of interest's pools do not overlap between rows and columns. Thus, the probability that the sample tests positive under the square array is
$P(Y_1,Y_2,Z|V=0) = P(Y_1|V=0) P(Y_2|V=0) P(Z|V=0)$ which is the same as in the alternate model with independent errors.
\qed
\end{proof}

\section{Derivation for the Linear Array Group Testing}
\label{Appendix B}
In the linear array testing, suppose the total number of people to test is $N$ and the desired group size is $n$. Note that the corresponding pool size is also $n$. We will form $ \lceil \frac{N}{n} \rceil$ linear array groups and perform group testing on each group. Note that if $N$ is not divisible by $n$, $\lfloor \frac{N}{n} \rfloor$ of the groups are of size $n$ and the last group is of size $N - \lfloor \frac{N}{n} \rfloor n$.
\subsection{Expected Number of Tests}
\label{Appendix A1}
The expected number of follow-up individual tests for a single pool of size $n$ is
\begin{gather*}
\begin{split}
m_{P}^{L}(n)  &= n  \sum_{d=0}^n  \bigg(1-\gamma(n,d)\bigg )\tbinom{n}{d} (1-p)^{n-d}(p)^d.
\end{split}
\end{gather*}

The total expected number of tests for the linear array testing of pool size $n$ is
\begin{gather*}
\begin{split}
M_{P}^{L}(N,n)  = \lceil \frac{N}{n} \rceil + \lfloor \frac{N}{n} \rfloor  m_{P}^{L}(n)  + m_{P}^{L}(N - \lfloor \frac{N}{n} \rfloor n).
\end{split}
\end{gather*}

Notice that $m_{P}^L\left(N-\lfloor \frac{N}{n}\rfloor n \right)$ is finite. The expected number of tests per person for the linear array testing of pool size $n$ for large $N$ is
\begin{gather*}
\begin{split}
M^{L}(n) &= \lim_{N \rightarrow \infty} \frac{1}{N} M_{P}^{L}(N,n) 
\\&=\frac{1}{n} + \sum_{d=0}^n  \bigg(1-\gamma(n,d)\bigg )\tbinom{n}{d} (1-p)^{n-d}(p)^d.
\end{split}
\end{gather*}

\subsection{Expected Number of False Negatives}
\label{Appendix A2}
The expected number of false negatives resulted from the group for a single linear array pool of size $n$ is
\begin{gather*}
\begin{split}
f_{G}^{L}(n)  &=  \sum_{d=1}^n d \gamma(n,d )\tbinom{n}{d} (1-p)^{n-d}(p)^d.
\end{split}
\end{gather*}
The expected number of false negatives for the follow-up individual tests for a single pool of size $n$ is 
\begin{gather*}
\begin{split}
f_{I}^{L}(n)&= \sum_{d=1}^n \eta(n,d)d \tbinom{n}{d} (1-p)^{n-d}(p)^d.
\end{split}
\end{gather*}

The expected number of false negatives for a single pool of size $n$ is
\begin{gather*}
\begin{split}
f_{P}^{L}(n) &= f_{I}^{L}(n) +f_{G}^{L}(n) 
\\ &= \sum_{d=1}^n \left(\eta(n,d)+\gamma(n,d)\right)d \tbinom{n}{d} (1-p)^{n-d}(p)^d.
\end{split}
\end{gather*}

The total expected number of false negatives for pool of size $n$ is
\begin{gather*}
\begin{split}
F_{P}^{L}(N,n) &= \lfloor \frac{N}{n} \rfloor f_{P}^{L}(n)  + f_{P}^{L}(N - \lfloor \frac{N}{n} \rfloor n).
\end{split}
\end{gather*}

Notice that $f_{P}^L\left(N-\lfloor \frac{N}{n}\rfloor n \right)$ is finite. The expected number of false negatives per person for a single pool of size $n$ for large $N$ is

\begin{gather*}
\begin{split}
F^{L}(n) &= \lim_{N \rightarrow \infty} \frac{1}{N} F_{P}^{L}(N,n) 
\\&=\sum_{d=1}^n \left(\eta(n,d)+\gamma(n,d)\right) \tbinom{n-1}{d-1} (1-p)^{n-d}(p)^d.
\end{split}
\end{gather*}

\subsection{Expected Number of False Positives}
\label{Appendix A3}
The expected number of false positives for a single pool of size $n$ is
\begin{gather*}
\begin{split}
\widetilde{f}_{P}^{L}(n) &= q\sum_{d=0}^{n-1} (1-\gamma(n,d))(n-d) \tbinom{n}{d} (1-p)^{n-d}(p)^d.
\end{split}
\end{gather*}

The expected total number of false positives, in $\lfloor \frac{N}{n}\rfloor$ arrays of size $n$ and one array of size $(N-\lfloor \frac{N}{n}\rfloor n )$, is
\begin{gather*}
\begin{split}
\widetilde{F}_{P}^L(N,n) = \lfloor \frac{N}{n}\rfloor \widetilde{f}_{P}^L(n) + \widetilde{f}_{P}^L\left(N-\lfloor \frac{N}{n}\rfloor n \right).
\end{split}
\end{gather*}

Notice that $\widetilde{f}_{P}^L\left(N-\lfloor \frac{N}{n}\rfloor n \right)$ is finite. The expected number of false positives per person for the linear array testing of pool size $n$ for large $N$ is
\begin{gather*}
\begin{split}
\widetilde{F}^{L}(n) &= \lim_{N \rightarrow \infty} \frac{1}{N} \widetilde{F}_{P}^{L}(N,n) 
\\&= q(1-p)\sum_{d=0}^{n-1} (1-\gamma(n,d)) \tbinom{n-1}{d} (1-p)^{n-d-1}(p)^d .
\end{split}
\end{gather*}
\section{Derivation for the Square Array Group Testing}
\label{Appendix C}
In square array testing, suppose the total number of people to test is $N$ and we consider the square array of size $n \times n$. Pools of size $n$ are created from each row and each column. We will form $\lfloor \frac{N}{n^2} \rfloor$ square arrays of size $n \times n$. Note that if $N$ is not divisible by $n^2$, for the remaining $N - \lfloor \frac{N}{n^2} \rfloor n^2$ people, we will fit the remaining $N - \lfloor \frac{N}{n^2} \rfloor n^2$ people into an incomplete matrix with row pool size $n$ and column pool size less than $n$.
\subsection{Expected Number of Follow-up Individual Tests}
\label{Appendix B1}
The expected number of follow-up individual tests for a single group of size $n \times n$ is
\begin{gather*}
\begin{split}
m_{P}^{S}(n)  &= n^2 \Bigg( p\sum_{d_1=1}^{n} \sum_{d_2=1}^{n} \theta(n,d_1,d_2)  \tbinom{n-1}{d_1-1}\tbinom{n-1}{d_2-1}p^{d_1+d_2-2}(1-p)^{2n-d_1-d_2}  \\ &\ \ \ \ \ \ \ \ + (1-p) \left( \sum_{d=0}^{n-1} (1-\gamma(n,d))\tbinom{n-1}{d}(p)^d(1-p)^{n-1-d} \right)^2 \Bigg).
\end{split}
\end{gather*}	

For the total N subjects, we need $\lfloor \frac{N}{n^2} \rfloor$ square array tests, and an incomplete square array test. Denoted by $m_R(N,n)$ the expected number of tests for the remaining group, including the number of pooled tests and the number of follow-up individual tests, $0 \leq m_R(N,n) \leq 2n + n^2$. Therefore, the expected total number of tests for the square array of group size $n \times n$ is
\begin{gather*}
\begin{split}
M_{P}^{S}(N,n) = \lfloor \frac{N}{n^2}  \rfloor \bigg(m_{P}^{S}(n)  + 2n\bigg) + m_R(N,n).
\end{split}
\end{gather*}

The expected number of tests per person for the square array testing of group size $n \times n$ for large $N$ is

\begin{gather*}
\begin{split}
M^{S}(n) &= \lim_{N \rightarrow \infty} \frac{1}{N} M_{P}^{S}(N,n) 
\\&= \frac{2}{n} +p\sum_{d_1=1}^{n} \sum_{d_2=1}^{n} \theta(n,d_1,d_2)  \tbinom{n-1}{d_1-1}\tbinom{n-1}{d_2-1}p^{d_1+d_2-2}(1-p)^{2n-d_1-d_2}  \\ &\ \ \ \ \ \ \ \ + (1-p) \left( \sum_{d=0}^{n-1} (1-\gamma(n,d))\tbinom{n-1}{d}(p)^d(1-p)^{n-1-d} \right)^2.
\end{split}
\end{gather*}

\subsection{Expected Number of False Negatives}
\label{Appendix B2}
The expected number of false negatives resulted from the group for a single square array of group size $n \times n$ is
\begin{gather*}
\begin{split}
f_{G}^{S}(n)  &= n^2 p \left(1 - \sum_{d_1=1}^{n} \sum_{d_2=1}^{n} \theta(n,d_1,d_2)  \tbinom{n-1}{d_1-1}\tbinom{n-1}{d_2-1}p^{d_1+d_2-2}(1-p)^{2n-d_1-d_2} \right).
\end{split}
\end{gather*}
The expected number of false negatives for the follow-up individual tests for a single square array group of size $n \times n$ is 
\begin{gather*}
\begin{split}
f_{I}^{S}(n) &= n^2 \sum_{d_1=1}^{n} \sum_{d_2=1}^{n}\psi(n,d_1,d_2) \tbinom{n-1}{d_1-1}\tbinom{n-1}{d_2-1}p^{d_1+d_2-1}(1-p)^{2n-d_1-d_2}.
\end{split}
\end{gather*}

The expected number of false negatives for a single square array group of size $n \times n$ is
\begin{gather*}
\begin{split}
f_{P}^{S}(n) &= f_{I}^{S}(n) +f_{G}^{S}(n) 
\\&= n^2p \bigg(1 - \sum_{d_1=1}^{n} \sum_{d_2=1}^{n} (\theta(n,d_1,d_2) -\psi(n,d_1,d_2))\\ &\ \ \ \ \ \ \ \ \tbinom{n-1}{d_1-1}\tbinom{n-1}{d_2-1}p^{d_1+d_2-2}(1-p)^{2n-d_1-d_2} \bigg).
\end{split}
\end{gather*}

Denoted by $f_R(N,n)$ the expected number of false negatives for the remaining group, $0 \leq f_R(N,n) \leq n^2$. The total expected number of false negatives for square array test is
\begin{gather*}
\begin{split}
F_{P}^{S}(N,n) &= \lfloor \frac{N}{n^2} \rfloor f_{P}^{S}(n)  +  f_R(N,n).
\end{split}
\end{gather*}

The expected number of false negatives per person for a single group test of size $n \times n$ for large $N$ is
\begin{gather*}
\begin{split}
F^{S}(n) &= \lim_{N \rightarrow \infty} \frac{1}{N} F_{P}^{S}(N,n) 
\\&= p \bigg(1 - \sum_{d_1=1}^{n} \sum_{d_2=1}^{n} (\theta(n,d_1,d_2) -\psi(n,d_1,d_2))\\ &\ \ \ \ \ \ \ \ \tbinom{n-1}{d_1-1}\tbinom{n-1}{d_2-1}p^{d_1+d_2-2}(1-p)^{2n-d_1-d_2} \bigg).
\end{split}
\end{gather*}

\subsection{Expected Number of False Positives}
\label{Appendix B3}
The expected number of false positives for a single group of size $n \times n$ is
\begin{gather*}
\begin{split}
\widetilde{f}_{P}^{S}(n) &= n^2 q (1-p) \bigg(\sum_{d=0}^{n-1} (1-\gamma(n,d)) \tbinom{n-1}{d}p^{d}(1-p)^{n-1-d} \bigg)^2.
\end{split}
\end{gather*}

Denoted by $\widetilde{f}_R(N,n)$ the expected number of false negatives for the remaining group, $0 \leq \widetilde{f}_R(N,n) \leq n^2$. The expected total number of false positives, in $\lfloor \frac{N}{n^2}\rfloor$ square arrays of size $n\times n$ and remaining incomplete square array, is
\begin{gather*}
\begin{split}
\widetilde{F}^S(N,n) = \lfloor \frac{N}{n^2}\rfloor \widetilde{f}^S(n) + \widetilde{f}_R(N,n).
\end{split}
\end{gather*}

The expected number of false positives per person for the square array testing of group size $n \times n$ for large $N$ is
\begin{gather*}
\begin{split}
\widetilde{F}^{S}(n) &= \lim_{N \rightarrow \infty} \frac{1}{N} \widetilde{F}_{P}^{S}(N,n) 
\\&= q (1-p) \bigg(\sum_{d=0}^{n-1} (1-\gamma(n,d)) \tbinom{n-1}{d}p^{d}(1-p)^{n-1-d} \bigg)^2.
\end{split}
\end{gather*}
\section{Mixed Array Group Testing}
\label{Appendix D}
Suppose we have a total of $N$ samples to test on each day. However, for a given pool size $n$, we can not always fill the $N$ samples into matrices of $n \times n$. 
We could fill the remaining samples into smaller matrices but it costs a lot to change pool size dynamically during experiment in classical lab setting. Instead, we consider using a mixed method of linear array and square array. 

In this method, we conduct $\lfloor\frac{N}{n^2}\rfloor$ square array tests and for the remaining $N-n^2\lfloor\frac{N}{n^2}\rfloor$ samples, we will do $\left\lfloor\frac{N-n^2\lfloor\frac{N}{n^2}\rfloor}{n}\right\rfloor$ linear array tests of pool size $n$ and a linear array test of pool size $N-n\lfloor\frac{N}{n}\rfloor$ if $N-n\lfloor\frac{N}{n}\rfloor\neq0$. Therefore, the expected total number of tests for the mixed array of pool size $n$ is:
\begin{gather*}
\begin{split}
M_{P}(N,n)  = \lfloor \frac{N}{n^2} \rfloor \bigg( m_{P}^{S}(n) + 2n\bigg) + \left\lceil\frac{N-n^2\lfloor\frac{N}{n^2}\rfloor}{n}\right\rceil \\+ \left\lfloor\frac{N-n^2\lfloor\frac{N}{n^2}\rfloor}{n}\right\rfloor   m_{P}^{L}(n)  + m_{P}^{L}(N - \lfloor \frac{N}{n} \rfloor n).
\end{split}
\label{expected number of tests for mixed}
\end{gather*}

The total expected number of false negatives for the mixed array test is:
\begin{gather*}
\begin{split}
F_{P}(N,n) = \lfloor \frac{N}{n^2} \rfloor f_{P}^{S}(n) + \left\lfloor\frac{N-n^2\lfloor\frac{N}{n^2}\rfloor}{n}\right\rfloor f_{P}^{L}(n) \\+ f_{P}^{L}(N - \lfloor \frac{N}{n} \rfloor n).
\end{split}
\label{expected number of false negatives for mixed}
\end{gather*}

The total expected number of false positives for the mixed array test is:
\begin{gather*}
\begin{split}
\widetilde{F}_{P}(N,n) = \lfloor \frac{N}{n^2} \rfloor \widetilde{f}_{P}^{S}(n) + \left\lfloor\frac{N-n^2\lfloor\frac{N}{n^2}\rfloor}{n}\right\rfloor \widetilde{f}_{P}^{L}(n) \\+ \widetilde{f}_{P}^{L}(N - \lfloor \frac{N}{n} \rfloor n).
\end{split}
\label{expected number of false positives for mixed}
\end{gather*}

We actively monitor and update the prevalence rate in each testing cycle. Given the total number of samples to test $N$, the prevalence rate $p$ in the beginning of each testing cycle, a given level of testing capacity per person $C$, and a given level of false positive tolerance $\widetilde{C}$, we optimize the pool size as follows:

\begin{gather}
    \begin{aligned}
        & \underset{n \in \{1,2,\cdots,\Bar{n} \}}{\text{minimize}}
        & & F_{P}(N,n) \\
        & \text{subject to}
        & & M_{P}(N,n) \leq CN,
        & & \widetilde{F}_{P}(N,n) \leq \widetilde{C}N.
    \end{aligned}
    \label{opt: section 5}
\end{gather}

\bibliographystyle{spbasic}  
\bibliography{references}
\end{document}